\documentclass[12pt]{iopart}

\expandafter\let\csname equation*\endcsname\relax
\expandafter\let\csname endequation*\endcsname\relax
\usepackage{amsmath}
\usepackage{amssymb}
\usepackage{graphicx}
\usepackage{enumitem}

\eqnobysec

\usepackage{epsfig}

\usepackage{cite} 

\usepackage{xcolor}

\usepackage[utf8]{inputenc}

\usepackage{mathtools}
\usepackage{bbold}
\usepackage{physics}

\usepackage{hyperref}
\usepackage{upgreek}

\newcommand{\beq}[1]{\begin{equation}{\label{eq:#1}}}
	\newcommand{\eeq}{\end{equation}}

\begin{document}



	\title{General theory of perturbation of infinite resistor networks}

	\author{J\'ozsef Cserti$^1$ and Gyula D\'avid$^2$}
	
	\address{$^1$E\"otv\"os Lor\'and University, Department of Physics of Complex Systems,
		H-1117 Budapest, P\'azm\'any P\'eter s\'et\'any 1/A, Hungary}
	
	\address{$^2$E\"otv\"os Lor\'and University, Institute of Physics and Astronomy,
		H-1117 Budapest, P\'azm\'any P{\'e}ter s{\'e}t\'any 1/A, Hungary}

	\begin{abstract}
		
		The effective resistance between any two nodes in a perturbed resistor network is determined by removing multiple bonds from an infinite resistor lattice. 
		We have developed an efficient method for calculating the Green operator of the Laplacian for such perturbed networks, which is directly related to the two-point resistance. 
		Unlike the recursive techniques that remove bonds one at a time, our approach handles all bond modifications simultaneously.
		To demonstrate the versatility of our method, several analytical and numerical examples are presented. 
		In addition, we computed bond current distributions to gain deeper insight into the nature of resistor perturbations. 
		We emphasize that our method has a broad range of applications, including condensed matter physics describing the quantum mechanical effects of impurities in crystal lattices, recently emerging topoelectronics, the study of vibrations in spring networks, and problems involving random walks.

	\end{abstract}
	
	
	\maketitle

	\section{Introduction}
	\label{bev:sec}
	

	Electric circuit theory as a classical problem has been studied for over one century, and it goes back to Kirchhoff's classical paper~\cite{Kirchhoff_classic_paper_1847}.
	The resistance between two arbitrary nodes in an infinite square lattice of identical resistors was first calculated by van der Pol and Bremmer~\cite{vanderPol}. This problem was later studied for the square lattice by Venezian~\cite{Venezian}, and for the square, triangular, and honeycomb lattices by Atkinson and Steenwijk~\cite{Atkinson_Steenwijk_10.1119/1.19311}. 
	In Doyle and Snell's book~\cite{Doyle}, which includes many intriguing results and relevant references, as well as Lovasz's paper~\cite{Lovasz_random-walk:cikk}, the relationship between random walks and electrical networks is explored. 
	This issue is also pertinent to the first-passage processes that were studied by Redner in his book~\cite{Redner:book}.
	
	In light of Kirkpatrick's seminal work~\cite{Kirkpatrick_RevModPhys.45.574}, a novel approach for calculating the effective resistance employing the Green operator was developed 
	in reference~\cite{sajat_perfect_AJP_10.1119/1.1285881}.
	Subsequently, the Green operator method was extended to calculate the resistance between two arbitrary nodes on any infinite periodic tiling of space~\cite{sajat_tiling_2011}.
	The application of Green operators to calculate electrical resistance has inspired numerous subsequent studies and developments. This approach has proven to be a powerful tool, enabling researchers to determine resistances in various lattice structures and configurations~\cite{Asad2013_FCC,Asad_square_Capacitor_2005,Asad_capacitor_3D_cubic_2013}.
	
	The problem of finite resistor networks is also of significant interest in circuit theory.
	Recently, Wu developed a theory to calculate the resistance between arbitrary nodes in finite resistor lattices~\cite{Wu_R_eigenv_2004}, while Tzeng and Wu extended this to impedance networks~\cite{Wu_Tzeng_finite_impedance:cikk}. 
	Essam and Wu determined the corner-to-corner resistance and its asymptotic expansion for free boundary conditions~\cite{Essam_Wu_corner_corner:cikk}, 
	Izmailian and Huang calculated this resistance 
	for other boundary conditions~\cite{PhysRevE.82.011125}.
	
	The perturbation of a perfect resistor network was studied by Kirkpatrick~\cite{Kirkpatrick_RevModPhys.45.574}, focusing on networks where resistors are randomly removed. 
	Inspired by this study, the Green operator of the resistor network was employed 
	in reference~\cite{sajat_perturbed_AJP_10.1119/1.1419104}.
	It was demonstrated that if a single resistor is removed from infinite perfect lattice, then the calculation of the Green operator for a perturbed network can be traced back to the Green operator of 
	the perfect network, utilizing the Sherman--Morrison formula~\cite{Sherman_Morrison_10.1214:cikk,Numerical_Recipes_3rd_10.5555:book}.
	This work has inspired numerous studies investigating different lattice structures~\cite{Asad_3d_cubic_1bonnd_removed_2004,Asad_square_perturbed_1bond_2006,Owaidat_perturb_substitutional_sq_2010,Owaidat_perturbed_tiling_2016}.
	The Sherman--Morrison formula can be applied recursively when removing more than one resistor. 
	The case of two removed resistors has been examined in references~\cite{Asad_two_bonds_2005,Asad_3d_cubic_2removed_bond_2008}.
	
	However, if more than two resistors are removed, the analytical calculations become less transparent and increasingly challenging to handle analytically. The subsequent application of single-bond perturbation in the case of a finite network was established by Bhattacharjee and Ramola~\cite{random_R_2023}. 
	Nonetheless, even for the case of four removed resistors, the resulting formulas are highly complex.
	
	The drawback of the recursive application of the Sherman–Morrison formula is that, at each step, the updated inverse matrix needs to be stored. This not only leads to the accumulation of numerical errors but also makes the method analytically rather difficult.
	This problem can be circumvented using the Woodbury matrix identity~\cite{Woodbury:cikk,Woodbury_Guttman_1177730946:cikk,Numerical_Recipes_3rd_10.5555:book}.  
	When a matrix is modified by a matrix of smaller rank, then its inverse can be obtained in terms of the inverse of the original matrix. 
	If both $\mathbf{A}$ and $\mathbf{C}^{-1}+\mathbf{V} \mathbf{A}^{-1}\mathbf{U}$ are invertible, then the Woodbury matrix identity is 
	\begin{align}
		\label{Woodbury:eq}
		{\left(\mathbf{A}+\mathbf{U}\mathbf{C}\mathbf{V}  \right)}^{-1}	
		&=
		\mathbf{A}^{-1}-\mathbf{A}^{-1} \mathbf{U} {\left(\mathbf{C}^{-1}+\mathbf{V} \mathbf{A}^{-1}\mathbf{U} \right)}^{-1} \mathbf{V} \mathbf{A}^{-1},
	\end{align}
	where $\mathbf{A}$ is $n\times n$, 
	$\mathbf{C}$ is $k\times k$, $\mathbf{U}$ is $n\times k$, and $\mathbf{V}$ is $k\times n$ matrices, $k\le n$.  
	The history of this formula is presented, and various applications to statistics, resistor networks, structural analysis, asymptotic analysis, optimization, and partial
	differential equations are discussed by Hager~\cite{Woodbury_Hager:cikk}. 
	Most of the power-system network problems involve the solution of very sparse linear equations. Then the perturbed sparse matrices corresponding to removal, addition, and splitting of nodes can be treated by the Woodbury matrix identity~\cite{Alsac_mod_network_4112038:cikk}.
	Other applications model how lightning flashes spread in a thundercloud~\cite{HAGER_villam_1989193:cikk} and study the breakdown properties of the random-fuse network~\cite{PhysRevB.36.367}.
	In circuit simulations, which involve numerous computations of the system response, using the Woodbury matrix identity allows one to avoid computationally expensive recalculations of the system response when system parameters change~\cite{Haley_1457374:cikk,dangling_large_R-network_5356296:cikk}.
	We should stress that in these examples, the networks involve a finite number of elements, that is, all matrices in the identity (\ref{Woodbury:eq}) are of finite size.
	
\begin{figure}[hbt]
		\centering
		\includegraphics[scale=0.6]{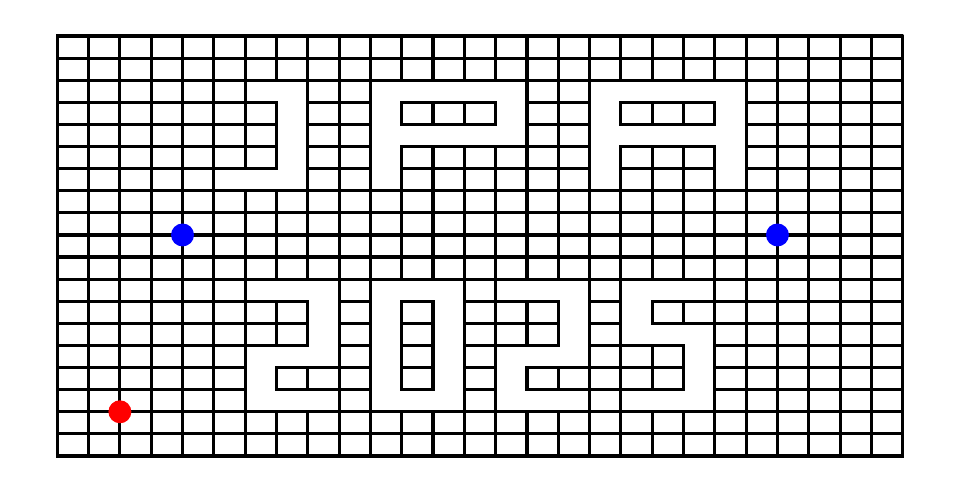}
		\caption{In an infinite square resistor network 
			$N=84$ bonds are removed to form a string of \textit{JPA  2025}.
			The red dot is the origin of the coordinate system.
			The resistance is calculated between sites 
			$\mathbf{r}_i =\left(2,8\right)$ and 
			$\mathbf{r}_j=\left(21,8\right)$ (blue dots). 
			\label{JPA:fig}}
	\end{figure}

	In what follows, we present a powerful formalism to calculate the effective resistance between two arbitrary lattice points in an \textit{infinite} perturbed resistor network. 
	To this end, we employ the Woodbury matrix identity to calculate the Green operator and the two-point resistance of a perturbed resistor lattice in which several resistors are replaced or removed.
	Our method can be used for both finite and infinite resistor networks.
	To demonstrate the effectiveness of our method, we calculate the two-point resistance in a perturbation of an infinite square resistor network, as shown in figure~\ref{JPA:fig}.
	This example represents a nontrivial perturbation of the infinite square resistor network, where a string \textit{JPA 2025}  containing numbers and letters results in the removal of 84 resistors.
	As shown below, we find that the effective resistance 
	(in units of $R$) between the sites 
	$\mathbf{r}_i =\left(2,8\right)$ and 
	$\mathbf{r}_j=\left(21,8\right) $ (the two blue dots in the figure) is 
	$R(r_i,r_j) = 1.6645...$ (for details see section~\ref{JPA_2025:sec}). 
	For comparison,  the resistance for perfect lattice is $R_0(r_i,r_j) = 1.45186...$. 
	
	The rest of the paper is organized as follows. In section~\ref{perfect_reiew_theory:sec}  the
	perfect lattice is reviewed following the approach in references~\cite{sajat_perfect_AJP_10.1119/1.1285881,sajat_perturbed_AJP_10.1119/1.1419104} using Dirac notation. 
	In section~\ref{gen_theory:sec}, we develop the Green operator of the perturbed resistor lattice. 
	In sections~\ref{anal_examples:sec} and~\ref{num_examples:sec}  we present a few analytical and numerical examples, respectively. 
	Conclusions are given in section~\ref{summary:sec}. 
	In~\ref{Green:app} the derivation of the Green operator is presented for infinite lattices.
	In~\ref{G_quantum:app} we highlight the relation between the energy‑dependent Green operator in quantum mechanics and the Green operator used for resistor networks.
	In~\ref{G_finite:app} the derivation of the Green operator is presented for finite lattices. 
	Finally, in~\ref{Hole_2_missing_nodes:app} and~\ref{3szog_hexagon:app}, details of calculations for hole-type perturbations in square and triangular lattices are presented, respectively. 
	
	\section{General theory of resistance networks }
	\label{perfect_reiew_theory:sec}

	Consider a general resistor network with an arbitrary structure, which may be either finite or infinite.
	The network consists of nodes labeled $i$, each located at a position vector $\mathbf{r}_i$. 
	Nodes are connected by resistors between node $i$ and $j$ (or more generally, complex impedances) characterized by a conductance $\gamma_{ij}=\gamma_{ji}$ (reciprocal of resistance). 
	It is useful to introduce dimensionless conductance $c_{ij}=R \,\gamma_{ij}$, where $R$ is a reference resistance.
	If two nodes are not connected, their corresponding conductance is zero, i.e., $c_{ij}=0$. 
	The diagonal elements $c_{ii}=0$ are not defined.
	Assume that the potential is $V_i$ at node $i$. 
	Note that these potentials are not uniquely determined because adding a constant value $V_0$ to all $V_i$ 
	does not affect the currents flowing through the resistors. 
	As we shall see later, this fact plays a crucial role in our theory.
	
	The current entering (or exiting) the network from outside the network at node $i$ is denoted by $I_i$ (as a convention, their values are positive if the current enters the network).
	The current conservation implies that $\sum_i I_i=0$.
	From Kirchhoff’s and Ohm's laws, it follows $\sum_j  c_{ij}\,(V_i-V_j) =R\,I_i$~\cite{Wu_R_eigenv_2004}.
	This equation can be recast in matrix form:
	\begin{subequations}
		\label{alapmx:eq}
		\begin{align}
			\mathbf{L}\,\mathbf{V} &=-R\,\mathbf{I}, \quad \textrm{where}\\
			L_{ij} &=\,(1-\delta_{ij})\, c_{ij}- z_i\,\delta_{ij}.
		\end{align}
	\end{subequations}
	where $z_i=\sum_{j\ne i} c_{ij}$, and $\mathbf{V}$ and $\mathbf{I}$ are vectors with components $V_i$ and $I_i$, respectively.
	
	Upon treating the infinite lattice, it is convenient to employ the bra-ket formalism, which is widely used in quantum mechanics. 
	Moreover, this approach was first introduced to the percolation problem by Kirkpatrick in reference~\cite{Kirkpatrick_RevModPhys.45.574}, 
	and it subsequently was used in reference~\cite{sajat_perfect_AJP_10.1119/1.1285881}.
	
	To this end, we introduce two vectors, $| V \rangle = \sum_i \ket i V_i  $ and 
	$| I \rangle= \sum_i \ket i I_i $ formed from the potentials $V_i = V(\mathbf{r}_i)$ and the currents $I_i = I(\mathbf{r}_i)$ at sites $\mathbf{r}_i$.
	Then, the Laplace operator can be written as  
	$\hat{L}=\sum_{i,j} \ket{i} L_{ij} \bra{j}$. 
	We shall call this form the \textit{site representation} of the Laplacian.
	The Laplace operator in this form is a linear combination of the dyadic products of vectors 
	corresponding to the sites. 
	Here, it is assumed that the ket vectors $\ket i$, associated with the site ${\bf r}_i$, form a complete orthonormal set in a finite or infinite dimensional linear space of sites, that is, $\langle i | j\rangle =\delta_{ij}$ 
	and $\sum_i \, | i \rangle \langle i | = \hat{\mathbb{1}}$, where $\delta_{ij}$ is the Kronecker delta symbol 
	and $\hat{\mathbb{1}}$ is the identity operator of the linear space of sites.
	Note that here $i$ and $j$ run over all the grid points of the infinite lattice.
	Then, equation~(\ref{alapmx:eq}) takes the operator form $\hat{L}\,\ket{V}=-R\,\ket{I}$, where
	\begin{align}
		\label{diadok:eq}
		\hat{L} &= \sum_{i,j} \ket{i} \Bigl [
		\left(1-\delta_{ij}\right) \, c_{ij}-\delta_{ij}\, \sum_{k\ne i} c_{ik} \Bigr ] \bra{j} 
		=\,-\frac{1}{2}\, \sum_{i,j} \bigl (\ket{j}-\ket{i} \bigr)\,c_{ij}\,\bigl (\bra{j}-\bra{i} \bigr).
	\end{align}
	In the last step, we collected separately the terms that contain the coefficients $c_{ij}$ related to the corresponding edge in the Laplace operator, and the factor $1/2$ takes into account the double summation over the lattice sites.

	We now define the ket vector $|b_k \rangle $ associated with the bond $k$ in the lattice as 
	\begin{align}
		\label{b_ved1_def:e}
		|b_k \rangle &= |e_k \rangle -|s_k\rangle, 
	\end{align} 
	where $|s_k \rangle$ and $|e_k \rangle$
	correspond to the sites 
	representing the start and end points of the bond $k$, respectively.
	With this notation, the Laplace operator in equation~(\ref{diadok:eq}) can be further simplified as 
	\begin{align}
		\hat{L} &=\,-\sum_{k} \ket{b_k} \beta_k \bra{b_k}, 
		\label{Ldiad:eq}
	\end{align}
	where $\beta_k$ denotes the normalized conductance of bond $k$, i.e., 
	$\beta_k = c_{e_k s_k}$.
	The Laplace operator becomes a sum of dyadic products corresponding to each bond. 
	We shall refer to the Laplace operator in this form as the \textit{bond representation} of the Laplacian.
	
	To determine the two-point resistance in a resistor network, it is useful to introduce the inverse of the Laplace operator 
	(provided it exists), commonly referred to in the literature as the lattice Green operator: 
	\begin{align}
		\hat{L}\hat{G}=-\hat{\mathbb{1}}.
		\label{Green:eq}
	\end{align}
	Knowing the Green operator for a given resistor network, various physical quantities can be calculated, such as the two-point resistance.    
	Namely, the two-point resistance between the sites $\mathbf{r}_i$ and $\mathbf{r}_j$
	can be expressed in terms of the matrix elements of the Green operator as follows~\cite{Kirkpatrick_RevModPhys.45.574,sajat_perturbed_AJP_10.1119/1.1419104,sajat_perfect_AJP_10.1119/1.1285881,sajat_tiling_2011}: 
	\begin{align}
		R_{ij} &=  R \,\bigl [\bra{i} \hat{G} \ket{i}  + \bra{j} \hat{G} \ket{j}  -2\,\bra{i} \hat{G} \ket{j} \bigr ]. 
		\label{Rij:eq}
	\end{align}
	Note that swapping $\ket{i}$ and $\ket{j}$, which correspond to the two sites, does not change the resistance, i.e. $R_{ij} = R_{ji}$.
	
	\subsection{Perfect infinite lattices}
	\label{perfect:sec}
	
	A special subclass of resistor networks is the regular infinite network. 
	In this case, the geometry of the lattice forms a $ d$-dimensional regular grid, with identical resistors $R$ located on each edge. 
	The reference resistor, of course, can be $R$.
	All lattice points are specified by position vectors ${\bf r}$ given in the form
	\begin{align}
		\label{r_points_d_dim:eq}
		{\bf r} &=l_1\mathbf{a}_1 +l_2\mathbf{a}_2 + \cdots + l_d\mathbf{a}_d,
	\end{align}
	where $\mathbf{a}_1,\mathbf{a}_2,\cdots,\mathbf{a}_d$ are independent
	primitive translation vectors, and $l_1, l_2,\cdots,l_d$ range through all integer values.
	
	In this regular resistor network, the Laplace operator is significantly simplified. 
	The matrix elements $c_{ij}$ will be either $1$ or $0$, depending on whether the $i$th and $j$th 
	nodes are connected or not, i.e., 
	whether they are neighbors of each other in the regular lattice. 
	Moreover, the coordination number is the same for each node.  
	This \textit{regular} Laplace operator will be denoted by $\hat{L}_0$ in the following~\cite{sajat_perfect_AJP_10.1119/1.1285881} 
	and is given by
	\begin{align}
		\hat{L}_0 &= \sum_{i,j}\, \ket i \,
		\bigl( \Delta_{ij} - z\, \delta_{ij} \bigr) \, \bra j,
		\label{L0-def:eq}
	\end{align}
	where $z$ is the number of neighbors (coordination number) of each
	lattice site (for example, $z=2d$ for a $d$-dimensional
	hypercubic lattice), and $\Delta_{kl}$ is unity if the sites
	${\bf r}_k$ and ${\bf r}_l$ are nearest neighbors and zero
	otherwise. The summation is taken over all lattice sites.
	This form of the Laplacian is the site representation of the operator $\hat{L}_0$. 
	
	Once again, we shall also present the Laplacian in its bond representation. 
	In this case, in equation~(\ref{Ldiad:eq}) for the nearest neighbors, the coefficient $\beta_k$ is $1$, otherwise it is $0$. 
	Then, for a regular lattice, we have
	\begin{align}
		\hat{L}_0 &= -\sum_{k} \ket{b_k} \bra{b_k},
		\label{L0diad:eq}
	\end{align}
	where the summation is over the bonds connecting the nearest neighbors. 
	Similarly, as in equation~(\ref{Green:eq}) we define the Green operator for the Laplace operator $\hat{L}_0$ as 
	\begin{align}
		\hat{L}_0 \hat{G}_0 &=-\hat{\mathbb{1}}.
		\label{G0-def}
	\end{align}
	This Green operator has been used to calculate the two-point resistance in square lattice 
	in references~\cite{sajat_perfect_AJP_10.1119/1.1285881,sajat_perturbed_AJP_10.1119/1.1419104}. 
	The properties of the lattice Green operator $\hat{G}_0$ and its explicit form in coordinate representation are given, e.g., in references~\cite{sajat_perfect_AJP_10.1119/1.1285881,Tony_Guttmann_2010,Mamode2021_Green_properties}.
	Owing to the translation symmetry of the perfect lattice, we have $ \bra i \hat{G}_0 \ket i = \bra j \hat{G}_0 \ket j$. 
	Therefore,  using equation~(\ref{Rij:eq}) the resistance between lattice points $\mathbf{r}_i$ and $\mathbf{r}_j$ in regular lattice is given by 
	\begin{align}
		R_0(r_i,r_j) &= 2 R \bigl [G_0(r_i,r_i) -G_0(r_i,r_j) \bigr ], 
		\label{R0:eq}
	\end{align}
	where $G_0(r_i,r_j) = \bra i \hat{G}_0 \ket j$ is the matrix element of $\hat{G}_0$.
	
	For convenience in table~\ref{R0:table} we present a few values of the resistances between sites ${\bf r}_i$ and ${\bf r}_j$ for a two-dimensional infinite square lattice.
	\begin{table}[hbt]
		\caption{\label{R0:table}
			Resistance $R_0(m,n)$ in an infinite square lattice in units of $R$.}
		\begin{indented}
			\item[]\begin{tabular}{@{}ccccc}
				\br
				$m \backslash n$ &	0 & 1 & 2 & 3 \\
				\mr
				0 &	0 & $\frac{1}{2}$ & $2-\frac{4}{\pi }$ &
				$\frac{17}{2}-\frac{24}{\pi }$ \\[2ex]
				1 &     $\frac{1}{2}$ & $\frac{2}{\pi }$ & $\frac{4}{\pi
				}-\frac{1}{2}$ & $\frac{46}{3 \pi }-4$ \\[2ex]
				2 &		$2-\frac{4}{\pi }$ & $\frac{4}{\pi }-\frac{1}{2}$ &
				$\frac{8}{3 \pi }$ & $\frac{1}{2}+\frac{4}{3 \pi }$ \\[2ex]
				3 &	$\frac{17}{2}-\frac{24}{\pi }$ & $\frac{46}{3 \pi }-4$ &
				$\frac{1}{2}+\frac{4}{3 \pi }$ & $\frac{46}{15 \pi }$ \\[2ex]
				\br
			\end{tabular}
		\end{indented}
	\end{table}
	Owing to lattice translation symmetry, for given sites $\mathbf{r}_i$ and $\mathbf{r}_j$ the indices $m$ and $n$ in the table and the resistance can be determined from the following equations:
	\begin{subequations}
		\label{mn_indices:def}
		\begin{align}
			\mathbf{r}_j-\mathbf{r}_i &=m\, \mathbf{a}_1 +n\, \mathbf{a}_2 ,
			\\[2ex]
			R_0(r_i,r_j) &\equiv 
			R_0(\left| m \right|, \left| n \right|). 
		\end{align}
	\end{subequations}
	More resistance values are tabulated in reference~\cite{Atkinson_Steenwijk_10.1119/1.19311}, and for arbitrary lattice sites, it can be obtained from the recurrence relations presented in reference~\cite{sajat_perfect_AJP_10.1119/1.1285881}.

	\section{General theory of the perturbation of infinite resistor networks}
	\label{gen_theory:sec}
	
	In the following, we first consider a perturbed network on an infinite resistor lattice containing no isolated regions; the network remains fully connected. 
	We then proceed to discuss the case of a disconnected network.
	
	When a single resistor is removed from the perfect network, the Laplacian for the perturbed lattice has been derived in references~\cite{Kirkpatrick_RevModPhys.45.574, sajat_perturbed_AJP_10.1119/1.1419104}.
	We now assume that in a perfect infinite lattice, in the $p$th bond, the resistance $R$ is replaced by a resistance $1/\gamma_p$.  Thus, the dimensionless conductance changes from $\beta_p=1$ to $\beta_p=R\,\gamma_p$.  In general,  $\beta_p$  can be different for each bond, 
	and $\beta_p = 0 $ if the corresponding bond is removed entirely.
	
	Owing to the change of the resistance in the $p$th bond the Laplace operator $\hat{L}_0$ 
	in equation~(\ref{L0diad:eq}) the coefficient of the dyadic product $\ket {b_p} \bra{b_p}$ 
	changes from $-1$ to $-\beta_p$. 
	This modification can be expressed by adding the dyadic product $\ket {b_p} g_p \bra{b_p}$ 
	to the operator $\hat{L}_0$, where $g_p = 1-\beta_p$. 
	If the $p$th bond is removed entirely, then $\beta_p = 0$ and thus $g_p=1$.
	When more than one resistors are replaced in the perfect lattice than the above procedure should be iterated. 
	Thus, if the resistances of $N$ number of  bonds are changed, the Laplace operator of the perturbed resistor network can be written as
	\begin{align}
		\label{L0+l1:eq}
		\hat{L} &= \hat{L}_0 +\hat{L}_1,  \quad \textrm{where} \quad 
		\hat{L}_1 = \sum_{p=1}^N | b_p \rangle g_p \langle b_p |. 
	\end{align}
	The Green operator   $\hat{G} = -\hat{L}^{-1}$ of the perturbed network is given by 
	\begin{subequations}
		\label{G:eq}
		\begin{align}
			\label{G_a:eq}
			\hat{G} &=
			-{\left(\hat{L}_0 
				+\sum_{p=1}^N | b_p \rangle g_p \langle b_p |\right)}^{-1} \equiv - {\left(\hat{L}_0 +|\alpha \rangle \mathbf{C} \langle \alpha | \right)}^{-1}, 
		\end{align}
		where we introduced the vector $|\alpha \rangle $ 
		and its adjoint $\langle \alpha |= {|\alpha \rangle }^\dag $, 
		and the diagonal matrix $\mathbf{C}$ as 
		\begin{align}	
			\label{alpha:def}
			|\alpha \rangle  &= \left(|b_1\rangle, |b_2\rangle,\dots,|b_N\rangle\right),\quad 
			\langle \alpha | = {|\alpha \rangle }^\dag
			= \begin{pmatrix}
				\langle b_1| & \\
				\langle b_2|& \\
				\vdots & \\
				\langle b_N|
			\end{pmatrix},  \\ 
			\label{C_def:eq}
			\mathbf{C} &= \mathrm{diag} (g_1,g_2, \dots, g_N) = \begin{pmatrix}   
				g_{1} & 0 & \cdots & 0 \\  
				0 & g_{2} & \cdots & 0 \\ 
				\vdots & \vdots & \ddots & \vdots \\  
				0 & 0 & \cdots & g_{N}   
			\end{pmatrix}  	.
		\end{align}
	\end{subequations}
	Here we adopted the notation $|\alpha \rangle$ used in Economou's book~\cite{economou2006green}.
	
	As we shall see, owing to the special form of $\hat{L}_1$, the inverse in equation~(\ref{G_a:eq}) can be calculated using the Woodbury identity~(\ref{Woodbury:eq}). 
	We find that the matrix element $G_{ij}=\langle i | \hat{G} | j\rangle $ of the Green operator   $\hat{G}$ is given by 
	(for details see~\ref{Green:app})
	\begin{subequations}
		\label{G_Woodbury:eq}
		\begin{align}
			\label{G_ri_rj:eq}
			G_{ij}  &=	\langle i |\,  \hat{G}_0 \, | j\rangle 
			+ \mathbf{U}^{\left(i\right)}  \,
			\mathbf{B}^{-1}  \, \mathbf{V}^{\left(j\right)} , 
			\quad \textrm{where} \quad
			\mathbf{B} = \mathbf{C}^{-1} 
			-  \langle \alpha | \, \hat{G}_0\,  |\alpha \rangle , 
			\quad \textrm{and}  \\[2ex]
			\label{ri_G0_a:eq}
			\mathbf{U}^{\left(i\right)} &= 
			\langle i |\, \hat{G}_0 \, |\alpha \rangle =	
			\langle i |  \,  \hat{G}_0 
			\left(|b_1\rangle, |b_2\rangle,\dots,|b_N\rangle\right)
			= \left(G_0(r_i,b_1), G_0(r_i,b_2),  \cdots ,  G_0(r_i,b_N)\right), \\
			\label{a_G0_rj:eq}
			\mathbf{V}^{\left(j\right)}  &=	
			\langle \alpha |  \,  \hat{G}_0 \, | j\rangle =
			{\left(
				\langle b_1|, \langle b_2|, \dots ,\langle b_N
				|\right)}^T	\,  \hat{G}_0\,  | j\rangle 
			= {\left[\mathbf{U}^{\left(j\right)}\right]}^T , 
		\end{align}
		and $T$ denotes the transpose of a vector. 
		Note that $\mathbf{U}^{\left(i\right)}=\langle r_i |\, G_0 \, |\alpha \rangle$ and 
		$\mathbf{V}^{\left(j\right)} = \langle \alpha |  \,  G_0 \, | r_j\rangle$ 
		are finite, $N$-component row and column vectors, respectively. 
		Finally,
		\begin{align}
			\label{a_G0_a:eq}
			\langle \alpha | \, \hat{G}_0\,  |\alpha \rangle
			&=
			\begin{pmatrix}
				\langle b_1| & \\
				\langle b_2|& \\
				\vdots & \\
				\langle b_N|
			\end{pmatrix} \hat{G}_0 \, 
			\left(|b_1\rangle, |b_2\rangle,\dots,|b_N\rangle\right) 
			= \begin{pmatrix}   
				G_0(b_1,b_1) & G_0(b_1,b_2) & \cdots & G_0(b_1,b_N) \\  
				G_0(b_2,b_1) & G_0(b_2,b_2) & \cdots & G_0(b_2,b_N) \\ 
				\vdots & \vdots & \ddots & \vdots \\  
				G_0(b_N,b_1) & G_0(b_N,b_2)   & \cdots & G_0(b_N,b_N)    
			\end{pmatrix}  	, 		
		\end{align}
		and $ G_0(b_p,b_q) = \langle b_p| \hat{G}_0  |b_q \rangle$ and $p,q = 1,2,\dots, N$. 
		Here $	\langle \alpha | \, \hat{G}_0\,  |\alpha \rangle$ is an $N \times N$ matrix with the matrix elements corresponding to the matrix elements of the unperturbed Green operator $\hat{G}_0$.
	\end{subequations}
	
	Using the ket vector $|\alpha \rangle$ equation~(\ref{G_ri_rj:eq}) can be written in a very compact  operator form:  
	\begin{align}
		\label{G_operator:eq}
		\hat{G} &= 
		\hat{G}_0 + \hat{G}_0 \, |\alpha \rangle \,
		{\left(\mathbf{C}^{-1} 
			-  \langle \alpha | \, \hat{G}_0\,  |\alpha \rangle\right)}^{-1}
		\, \langle \alpha |  \,  \hat{G}_0 . 
	\end{align}
	Equations~(\ref{G_Woodbury:eq}) and (\ref{G_operator:eq}) are one of the main results of this work. 
	Because the Woodbury identity may not be familiar to all readers, we present an alternative derivation of equation~(\ref{G_operator:eq}) in~\ref{G_quantum:app} that is more accessible to readers with a physics background.
	
	As can be seen, this expression of the Green operator has a very compact form. 
	When the operator $\hat{L}_1$ is only one dyadic product, i.e., only one bond is removed from the perfect lattice, 
	then equation~(\ref{G_operator:eq}) reduces to the Sherman--Morrison formula~\cite{sajat_perturbed_AJP_10.1119/1.1419104}. 
	In principle, the Sherman-Morrison formula can be used subsequently for more than one bond perturbation. 
	For the case of two bonds, see reference~\cite{Asad_two_bonds_2005,Asad_3d_cubic_2removed_bond_2008}. 
	In the general case, this recursive method is used in reference~\cite{random_R_2023}.   
	However, this method is not easy to handle analytically. 
	Even for the case of four removed resistors, the analytical expressions presented in that work are quite complex~\cite{random_R_2023}.  
	In contrast, our method uses a non-recursive algorithm that can easily be applied to more complex perturbations, even when the number of removed resistors is significantly larger than one.
	We emphasize that in our method, for an arbitrary but finite number of bonds, only a single finite-dimensional matrix inversion is required, whereas in the recursive method, one needs to calculate the actual Green operator at each step.
	As we shall show, in cases where only a few bonds are removed, the calculations can be done by hand, yielding analytical results for the resistance.
	
	Up to now, the result for the Green operator of a Laplace operator $\hat{L}_0$  modified by a sum of dyadic products (outer product) given by $\hat{L}_1$ is general, independent of the specific form of the operator $\hat{L}_0$. 
	In addition, it is important to note that our derivation and the results we obtained for the perturbed Green operator can be applied to finite lattice systems. 
	It is worth emphasizing that, e.g., in quantum mechanical calculations, using the bra-ket formalism, our approach can be utilized to study the role of impurities within the tight-binding approximation~\cite{economou2006green}, as the sum of dyadic products modifies the Hamiltonian of the system.

	In what follows, we apply this result to our resistance problem.
	The matrix elements of the matrix  
	$\langle \alpha | \, \hat{G}_0\,  |\alpha \rangle$ 
	in equation~(\ref{a_G0_a:eq}) can be expressed in terms of the effective resistance in the perfect lattice as: 
	\begin{subequations}
		\label{gu_vg:eq}
		\begin{align}
			\label{G_pq:eq}
			\langle \alpha | \, \hat{G}_0\,  |\alpha \rangle _{pq}&= 
			G_0(b_p,b_q)=\langle b_p| \hat{G}_0  |b_q \rangle = 
			(\langle e_p|- \langle s_p|)\, \hat{G}_0 \,  (|e_q \rangle -|s_q\rangle) 
			\nonumber \\  
			&= \langle e_p|\, \hat{G}_0 \, |e_q \rangle 
			- \langle s_p|\, \hat{G}_0 \, |e_q \rangle 
			- \langle e_p|\, \hat{G}_0 \, |s_q \rangle
			+  \langle s_p|\, \hat{G}_0 \, |s_q \rangle 
			\nonumber \\
			&= \frac{1}{2R}\, 
			\left[
			-R_0(e_p,e_q)
			+R_0(s_p,e_q)
			+R_0(e_p,s_q)
			-R_0(s_p,s_q)
			\right].
		\end{align}
		In the last step we applied equation~(\ref{R0:eq}). 
		This result is independent of the type of perfect lattice. 
		In particular, for an infinite square lattice, the two-point resistance depends only on 
		the two components $m$ and $n$ of the relative difference vector $\mathbf{r}_i - \mathbf{r}_j$, and can be obtained from table~\ref{R0:table} using equation~(\ref{mn_indices:def}).
		
		Similarly as in equation~(\ref{G_pq:eq}), the matrix elements
		$G_0(r_i,b_p) = \langle i |\, \hat{G}_0\, |b_p\rangle $ and 
		$G_0(b_q,r_j) = \langle b_q |\, \hat{G}_0\, | j\rangle$
		in equations~(\ref{ri_G0_a:eq}) and (\ref{a_G0_rj:eq}) can also be expressed 
		in terms of the effective resistances in a perfect lattice: 
		\begin{align}
			\label{ri_G0:eq}
			G_0(r_i,b_p) &= \langle i |\, \hat{G}_0\, |b_p\rangle 
			=
			\langle i |\,  \hat{G}_0 \,|e_p \rangle 
			- \langle i |\,  \hat{G}_0 \,|s_p \rangle 
			= 
			\frac{1}{2R}\, 
			\left[
			-R_0(r_i,e_p)
			+R_0(r_i,s_p)
			\right], \\
			G_0(b_q,r_j) &= 	\langle b_q |\, \hat{G}_0\, | j\rangle 
			\label{G0_rj:eq}
			=
			\langle e_q |\,  \hat{G}_0 \,|j \rangle 
			- \langle s_q |\,  \hat{G}_0 \,|j \rangle 
			= 
			\frac{1}{2R}\, 
			\left[
			-R_0(r_j,e_q)
			+R_0(r_j,s_q)
			\right],
		\end{align}
	\end{subequations} 
	where $p,q = 1,2, \dots, N$.
	Here, we used the fact that in the perfect lattice, interchanging the two sites $r_a$ and $r_b$, the resistance does not change, i.e.,
	$R_0(r_b,r_a) = R_0(r_a,r_b)$.
	
	The effective resistance in the perturbed resistor network between sites $\mathbf{r}_i$ and $\mathbf{r}_j$ is given by equation~(\ref{Rij:eq}).
	
	Note that, unlike in a perfect lattice, $G(r_i,r_i) \ne G(r_j,r_j)$ because translation symmetry is broken in the perturbed lattice. 
	However, for resistor networks, the Laplacian in site representation is a real symmetric matrix, i.e., 
	$L(r_i,r_j) = L(r_j,r_i)$. Thus, it follows that 
	the matrix form of the Green operator is also a real symmetric matrix, i.e., 
	$G(r_i,r_j) = G(r_j,r_i)$.
	If the network contains impedances (capacitors and/or inductors), then the matrix form of the Laplacian 
	remains symmetric, but its matrix elements are complex, and consequently the Green operator is also a complex symmetric (but not Hermitian) matrix~\cite{Wu_Tzeng_finite_impedance:cikk}.
	Even so, equation~(\ref{Rij:eq}) is still valid for calculating the two-point impedance. 
	
\begin{figure}[hbt]
		\centering
		\includegraphics[scale=0.65]{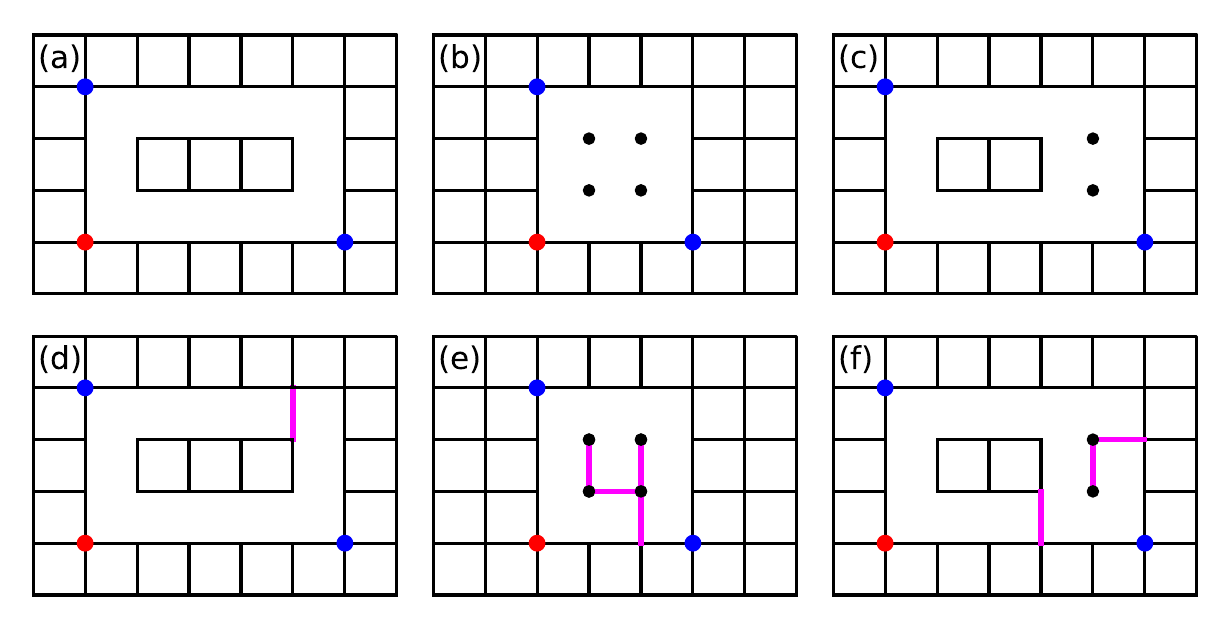}
		\caption{(a) Disconnected \textit{island}. 
			(b) A \textit{lake} created by four single-site islands (four isolated sites).
			(c) Extended island and two single-site islands (two isolated sites). 
			(d) A \textit{bridge} (magenta line) connects the island to one site of the perimeter of the infinite lattice.
			(e) Introducing \textit{dangling bonds}  (magenta lines) that pass through the four isolated sites.
			(f) A bridge connects the extended island, and dangling bonds pass through the two isolated sites.
			The resistance is calculated between sites $\mathbf{r}_i$ and $\mathbf{r}_j$ (blue dots). 
			The red dot is the origin of the coordinate system.
			\label{topological_issue:fig}
		}
	\end{figure}

	Now we turn to the case of disconnected networks. 
	A network can become disconnected when specific combinations of resistors are removed ($\gamma_p = 0$). 
	In such \textit{topological defects}, the network may contain either 
	an \textit{island}---a region separated from the rest of the network---or a \textit{lake}, which consists of one or more isolated sites (i.e., can be regarded as such islands that contain only a single site), as shown in figure~\ref{topological_issue:fig}(a) and (b), respectively.
	Moreover, there may be combinations of extended islands and isolated sites (an example is illustrated in figure~\ref{topological_issue:fig}(c)).
	
	For a given current distribution $| I \rangle$ in a disconnected lattice, the equation 
	$\hat{L} | V \rangle = -R  | I \rangle$ for the perturbed resistor network does not have a unique solution for the potentials $| V \rangle$ 
	at the lattice sites. 
	This is because the potentials can be shifted by the same values in each disconnected region, but with different values for each region.
	In each isolated region, independently shifted potential values define a distinct zero eigenvector.
	The operator $\hat{L}$ possesses multiple zero eigenvalues with linearly dependent eigenvectors. 
	Therefore, the inverse of the operator $\hat{L}$, and thus the Green operator, does not exist.
	Then, regarding the calculations of the two-point resistance, 
	the following cases should be considered:  
	\begin{enumerate}
		\item[a)] The sites $\mathbf{r}_i$ and $\mathbf{r}_j$, where the current enters and exits from the lattice, are on the same island. 
		In this case, the island can be regarded as a finite resistor network. 
		The calculation of the Green operator can then be performed as described in~\ref{G_finite:app}.
		
		\item[b)] The sites $\mathbf{r}_i$ and $\mathbf{r}_j$ 
		belong to two disconnected regions. For example, one site may be located on an island, while the other is situated outside this region, either in the infinite region or on another island.
		In this case, no current flows between the two sites, and the two-point resistance is infinite.
		\item[c)] Both sites $\mathbf{r}_i$ and $\mathbf{r}_j$ are in the infinite part of the resistor network. 
		
	\end{enumerate}
	
	In case c), each isolated island should be connected only by one \textit{bridge} to the infinite part of the network (see, e.g., figure~\ref{topological_issue:fig}(d) related to figure~\ref{topological_issue:fig}(a)). 
	For lakes, one should introduce \textit{dangling bonds} that pass through all the isolated sites in the lake and connect to the perimeter of the infinite grid at only one site (see, e.g., figure~\ref{topological_issue:fig}(e) related to figure~\ref{topological_issue:fig}(b)).
	Furthermore, in the case of more complex perturbations, such as the one shown 
	in figure~\ref{topological_issue:fig}(f), one should use both bridges and dangling bonds, 
	as illustrated in figure~\ref{topological_issue:fig}(c). 
	
	Since the bridge and/or the dangling bonds are attached to the infinite network at only a single site, no currents flow through these bonds. 
	After such modifications, the entire network becomes connected, that is, every site can be reached from any other site via connected bonds, but the current distribution remains unchanged. 
	Hence, the operator $\hat{L}$ of the perturbed lattice and consequently, the matrix $\mathbf{B}$ 
	in equation~(\ref{G_ri_rj:eq}) can be invertible.
	Moreover, the Green operator of the perturbed network exists, and it can be used in equation~(\ref{Rij:eq}) to obtain the two-point resistance.
	We will present examples for calculating the effective resistance for such topological defects later.
	
	Finally, using the matrix $\mathbf{B}$, and the vectors $\mathbf{U}^{\left(i\right)}$ and 
	$\mathbf{V}^{\left(j\right)}$  defined in equation~(\ref{G_Woodbury:eq}), the effective resistance between 
	$\mathbf{r}_i$ and $\mathbf{r}_j$ 
	in equation~(\ref{Rij:eq}) becomes
	\begin{align}
		\label{R_ij_perturbed:eq}
		R(r_i,r_j) &= 
		R_0(r_i,r_j) 
		+ R \left( 
		\mathbf{U}^{\left(i\right)} - \mathbf{U}^{\left(j\right)}
		\right)
		\mathbf{B}^{-1}
		\left( 
		\mathbf{V}^{\left(i\right)} - \mathbf{V}^{\left(j\right)}
		\right).
	\end{align}
	The resistance of the perturbed system is expressed in terms of the resistance of the corresponding perfect lattice. 
	We emphasize that the dimensions of the matrix and vectors involved in the above calculations are finite, specifically equal to the number of bonds removed from the perfect lattice.
	This fact ensures that the algorithm can be applied easily, analytically, and can be implemented straightforwardly in a programming language.
	In summary, the following steps are required to calculate the effective resistance between two arbitrary sites in the perturbed lattice: 
	
	\begin{enumerate}[label={\arabic*}.]
		
		\item Using equations~(\ref{G_ri_rj:eq}) and (\ref{G_pq:eq}) calculate the matrix elements of 
		$\langle \alpha | \, \hat{G}_0\,  |\alpha \rangle$.
		
		\item Calculate $\mathbf{C}^{-1} = \mathrm{diag} (\frac{1}{g_1},\frac{1}{g_2}, \dots, \frac{1}{g_N}) $  and the matrix $\mathbf{B}$ defined in equation~(\ref{G_ri_rj:eq})
		
		\item Calculate the inverse matrix $\mathbf{B}^{-1}$.
		
		\item Using equations~(\ref{ri_G0_a:eq}), (\ref{a_G0_rj:eq})  and (\ref{gu_vg:eq}) calculate the vectors $\mathbf{U}^{\left(i\right)}$,  
		$\mathbf{U}^{\left(j\right)}$, 
		$\mathbf{V}^{\left(i\right)}$ 
		and $\mathbf{V}^{\left(j\right)}$.

		\item The two-point resistance of the perturbed resistor network 
		can be obtained from equation~(\ref{R_ij_perturbed:eq}).
		
	\end{enumerate}
	Note that using equation~(\ref{G_ri_rj:eq}), one can also obtain the matrix elements $\hat{G}(r_i,r_j)$ of the Green operator for the perturbed lattice.
	We should emphasize that in the procedure outlined above, one needs to calculate only a finite number of matrix elements of the operator $\hat{G}_0$. 
	For instance, for a square lattice, the corresponding resistance can be calculated from the recurrence relations derived in reference~\cite{sajat_perfect_AJP_10.1119/1.1285881}.   
	
	Finally, to calculate the bond current distribution in the resistor networks, we give the expression for the current flowing through a given bond, which will be useful in the following examples as well. 
	To measure the resistance, it is assumed that a current $I_0$ 
	enters  from outside the lattice at the lattice point $\mathbf{r}_i$ and exits 
	at the lattice point $\mathbf{r}_j$. 
	Denote the two ends of a given bond by $\mathbf{r}_x$ and $\mathbf{r}_y$ and the conductance of this bond by $\gamma_{x,y}$. 
	Then the bond current flowing from $\mathbf{r}_x$  to $\mathbf{r}_y$  can be calculated from the potentials 
	between the two ends of the bond:
	$I(r_x,r_y|r_i,r_j) = 
	\gamma_{xy}\, \left[V(r_y) - V(r_x)\right]$
	and it becomes
	\begin{align}
		\label{current:eq}
		I(r_x,r_y|r_i,r_j) &=
		-\frac{I_0\, \gamma_{xy}}{2} \left[
		R(r_i,r_x) - R(r_i,r_y) -R(r_j,r_x) + R(r_j,r_y) 
		\right],
	\end{align}
	where $R(r_a,r_b)$ is the resistance between the lattice sites $\mathbf{r}_a$ and $\mathbf{r}_b$ either in the perfect or the perturbed lattice.   
	If the value of the current is negative, then the direction of the current flow is opposite.
	This expression allows us to calculate the 
	bond current distribution both in the perfect and perturbed lattice. 
	In the following, we demonstrate the efficiency of our general method.
	
	\section{Analytical examples  }
	\label{anal_examples:sec}
	
	First, we present a few analytical results to demonstrate how the resistance can be calculated in perturbed lattices of square and triangular resistor networks.
	The examples provided below show cases where obtaining analytical results for the resistance is not straightforward and transparent using the recursive application of the Sherman–Morrison formula.
	
	From now on, we consider only perturbations where the resistors are removed from the perfect resistor network.
	In this case, $\gamma_p =0 $ for all removed resistors, which implies $g_p = 1$ for all $p=1,2,\dots, N$ in equation~(\ref{C_def:eq}). 
	Hereafter, all results for the resistance are given in units of  $R$. 
	
	\subsection{Four bonds are removed}
	\label{4_bound:sec}

\begin{figure}[hbt]
		\centering
		\includegraphics[scale=0.5]{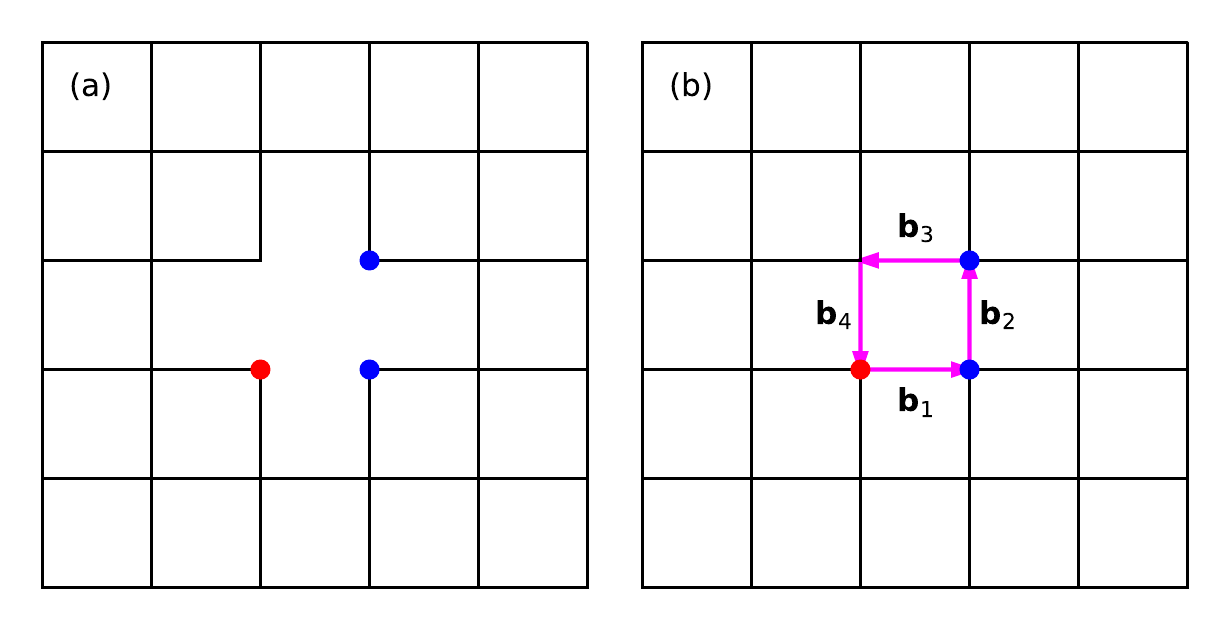}
		\caption{(a) Perturbation of an infinite square resistor network by removing four bonds. 
			(b) The bond vectors (magenta arrows) are 
			$\mathbf{b}_1, \mathbf{b}_2$, $\mathbf{b}_3$ and 
			$ \mathbf{b}_4$.
			The resistance is calculated between sites $\mathbf{r}_i$ and $\mathbf{r}_j$ (blue dots). 
			The red dot is the origin of the coordinate system.
			\label{4bonds_sq_A:fig}}
	\end{figure}

	Figure~\ref{4bonds_sq_A:fig}(a) shows the infinite square resistor network after removing four bonds. 
	Now, in this perturbed lattice, we calculate the resistance between lattice points 
	$\mathbf{r}_i =\left(1,0\right)$ and 
	$\mathbf{r}_j =\left(1,1\right)$, where the coordinates of lattice points are defined 
	in equation~(\ref{r_points_d_dim:eq}) for $d=2$ dimensions. 
	The coordinates of the four bond vectors shown in figure~\ref{4bonds_sq_A:fig}(b) are 
	$\mathbf{b}_l = \mathbf{e}_l -\mathbf{s}_l$, where $l=1,2,3,4$,  
	and $\mathbf{s}_1 = \left(0,0\right)$, 
	$\mathbf{e}_1 = \mathbf{s}_2 = \left(1,0\right) $, 
	$\mathbf{e}_2 = \mathbf{s}_3 = \left(1,1\right)$, 
	$\mathbf{e}_3 = \mathbf{s}_4 = \left(0,1\right)$,
	$\mathbf{e}_4 = \mathbf{s}_1$.
	In the present case, the origin of the coordinate system is at the starting point of the bond vector $\mathbf{b}_1$.
	Note that the results for the resistance are independent of the chosen origin of the coordinate system. 
	Furthermore, due to the symmetry of the perturbed lattice, the resistance remains identical when the endpoints of another removed bond are selected.
	
	First, we calculate the matrix 
	$\langle \alpha | \, \hat{G}_0\, |\alpha \rangle$. 
	To see how these calculations can be performed, we evaluate, for example, the matrix element $G_0(b_1,b_2)$. 
	From equation~(\ref{G_pq:eq}) and using the values of resistance for an infinite square resistor network given in table~\ref{R0:table}, we find that
	\begin{align}
		& \langle \alpha | \, \hat{G}_0\,  |\alpha \rangle_{12}
		= 	G_0(b_1,b_2) = 
		\frac{1}{2} 
		\left[
		-R_0(e_1,e_2)+R_0(s_1,e_2)+R_0(e_1,s_2)-R_0(s_1,s_2)  
		\right]  \nonumber \\[2ex]
		&=
		\frac{1}{2} 
		\left[
		-R_0((1,0),(1,1))+R_0((0,0),(1,1))
		+R_0((1,0),(1,0))-R_0((0,0),(1,0))
		\right] \nonumber \\[2ex]
		&= 
		\frac{1}{2} 
		\left[
		-2 R_0(0,1)+R_0(1,1)
		\right] = \frac{1}{2} \left(\frac{2}{\pi}-1\right) .
	\end{align}
	In the last but one step, we used equation~(\ref{mn_indices:def}) and the relation $R_0(r_j,r_i) = R_0(r_i,r_j)$.
	Similarly, the other matrix elements of $\langle \alpha | \, \hat{G}_0\,  |\alpha \rangle$ can be obtained 
	for the four bond vectors $\mathbf{b}_1,\mathbf{b}_2,\mathbf{b}_3$ and $\mathbf{b}_4$. 
	Then, we have
	\begin{align}
		\langle \alpha | \, \hat{G}_0\,  |\alpha \rangle
		&= \frac{1}{2}\, \begin{pmatrix}   
			1 & a & b & a\\  
			a & 1 & a & b \\ 
			b & a & 1 & a \\
			a & b & a & 1   
		\end{pmatrix},
	\end{align}
	where 
	$a= R_0(1,1)-2R_0(1,0)= \frac{2}{\pi}-1$ and 
	$b=2R_0(1,0)- 2 R_0(1,1)= 1-\frac{4}{\pi}$.
	In our case, $\mathbf{C} = \mathrm{diag} (1,1, 1, 1)$ and the matrix $\mathbf{B}$ 
	can be obtained from equation~(\ref{G_ri_rj:eq}).
	
	Now, using equations~(\ref{ri_G0_a:eq}), (\ref{a_G0_rj:eq}) and (\ref{gu_vg:eq}) 
	we calculate the vectors $\mathbf{U}^{\left(i\right)}$ 
	and $\mathbf{V}^{\left(j\right)}$.
	To illustrate the details, we evaluate, for instance, the third component of the vector $\mathbf{U}^{\left(i\right)}$. 
	Thus, for $\mathbf{r}_i =\left(1,0\right)$ we may write:
	\begin{align}
		\mathbf{U}^{\left(i\right)}_3 
		&= \langle i | \, \hat{G}_0\, | b_3 \rangle 
		= 	\frac{1}{2R}\, 
		\left[
		-R_0(r_i,e_3)
		+R_0(r_i,s_3)
		\right] \nonumber \\[2ex]
		&=
		\frac{1}{2} 
		\left[
		-R_0((1,0),(0,1))+R_0((1,0),(1,1))
		\right] = \frac{1}{2} 
		\left[
		-R_0(1,1)+R_0(0,1)
		\right] = \frac{b}{4}.
	\end{align} 
	Again, in the last but one step, we used equation~(\ref{mn_indices:def}) and the relation $R_0(r_j,r_i) = R_0(r_i,r_j)$.
	In a similar manner, one can calculate all components of the vectors 
	$\mathbf{U}^{\left(i\right)}$ 
	and $\mathbf{V}^{\left(j\right)}$.
	Finally, using equation~(\ref{R_ij_perturbed:eq}) the exact resistance of the perturbed resistor network 
	shown in figure~\ref{4bonds_sq_A:fig}(a) for two cases, namely between sites 
	$\mathbf{r}_i =\left(1,0\right)$ and 
	$\mathbf{r}_j =\left(1,1\right)$, and 
	$\mathbf{r}_i =\left(1,0\right)$ and 
	$\mathbf{r}_j =\left(0,1\right)$.
	Note that the matrix $\mathbf{B}$ and its inverse remain the same in both cases, so the inverse needs to be calculated only once.
	
\begin{figure}[hbt]
	\centering
	\includegraphics[scale=0.5]{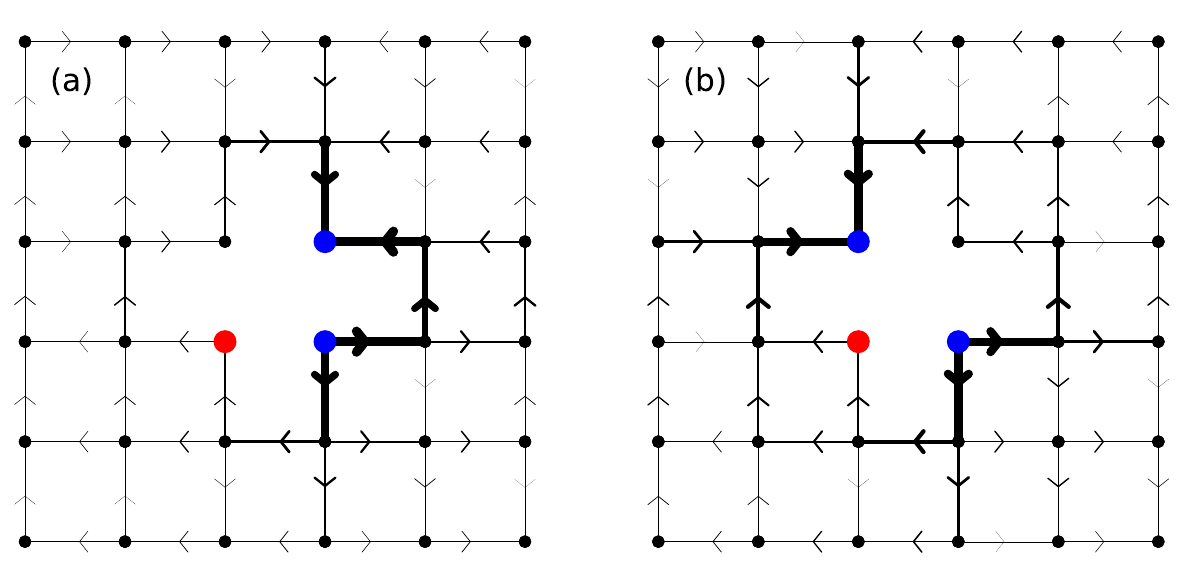}
	\caption{Bond current distribution  of the perturbed lattice 
		shown in figure~\ref{4bonds_sq_A:fig}(a)  
		for (a)
		$\mathbf{r}_i =\left(1,0\right)$ and 
		$\mathbf{r}_j =\left(1,1\right)$, and
		(b) $\mathbf{r}_i =\left(1,0\right)$ and 
		$\mathbf{r}_j =\left(0,1\right)$ (blue dots).	 
		The red dot is the origin of the coordinate system.
		The width of the arrow lines is proportional to the current value. 
		\label{4bonds_sq_C:fig}}
\end{figure}

	The results are summarized in table~\ref{4bonds_UV_R:table}.
	\begin{table}[hbt]
		\caption{\label{4bonds_UV_R:table}
			The vectors $\mathbf{U}^{\left(i\right)}$ and $\mathbf{V}^{\left(j\right)} $, and 
			resistances $R(r_i,r_j) $ and $R_0(r_i,r_j) $ (in units of $R$)
			in the perturbed and perfect square lattice, respectively.
		}
		\begin{indented}
			\item[]\begin{tabular}{@{}ccc}
				\br
				&	
				$\mathbf{r}_i =\left(1,0\right)$,  
				$\mathbf{r}_j =\left(1,1\right)$ 
				& $\mathbf{r}_i =\left(1,0\right)$,  
				$\mathbf{r}_j =\left(0,1\right)$  \\ 
				\mr
				$\mathbf{U}^{\left(i\right)}$ 
				&	 $\frac{1}{4}\, \left(1,-1,b,-b\right)$ 
				& $\frac{1}{4}\, \left(1,-1,b,-b\right) $  \\[2ex]
				$\mathbf{V}^{\left(j\right)} $ 
				&	$\frac{1}{4}\, {\left(-b,1,-1,b\right)}^T$ 
				& $\frac{1}{4}\, {\left(b,-b,1,-1\right)}^T$  \\[2ex]
				$R(r_i,r_j) $ &	$-\frac{12 - 6 \pi + \pi^2}{2\left(8-6\pi +\pi^2\right)} \approx 1.5409...$ 
				& $\frac{2}{\pi-2} \approx 1.7519...$  \\[2ex]
				$R_0(r_i,r_j) $ &	$\frac{1}{2}$ 
				& $\frac{2}{\pi} \approx 0.62666...$  \\[2ex]
				\br
			\end{tabular}
		\end{indented}
\end{table}
In figure~\ref{4bonds_sq_C:fig}, the bond current distribution, calculated from equation~(\ref{current:eq}), is plotted when the current enters and exits at the following lattice points: 
$\mathbf{r}_i =\left(1,0\right)$ and 
$\mathbf{r}_j =\left(1,1\right)$, 
and 
$\mathbf{r}_i =\left(1,0\right)$ and 
$\mathbf{r}_j =\left(0,1\right)$.

The resistance between the sites $\mathbf{r}_i = (1, 0)$ and $\mathbf{r}_j = (0, 1)$ can be evaluated using basic symmetry arguments. 
Between these two sites in the perfect lattice, the two-point resistance is $R_0(1,1)=2R/\pi$. 
Now consider the node at position $(1,1)$ in the perfect lattice. 
This node can be split into two distinct nodes: one connected to $(1,0)$ and $(0,1)$, 
and the other to $(2,1)$ and $(1,2)$.  
This transformation does not alter the effective resistance because the two split nodes are equipotential due to the symmetry of the system. 
A similar splitting can be applied to the node at $(0,0)$. 
After these modifications, the perfect lattice can be interpreted as a perturbed lattice 
with resistance $R_{\textrm{pert}}$, connected in parallel with two resistors, 
each of resistance $2R$.
Based on this construction, one can write 
$\frac{1}{R_0(1,1)}=\frac{1}{R_\textrm{pert}}+\frac{1}{2R}+\frac{1}{2R} $, and hence 
$R_\textrm{pert}=\frac{2}{\pi-2}R$. 

\subsection{A lake, as a topological defect }
\label{topo_holes:sec}

\begin{figure}[hbt]
	\centering 
	\includegraphics[scale=0.7]{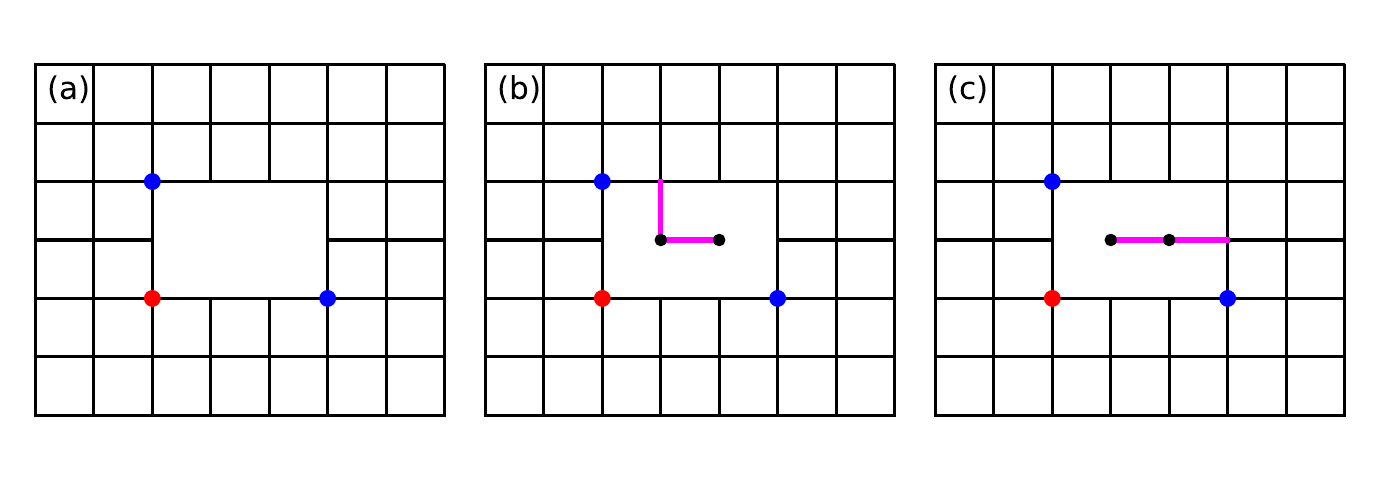}
	\caption{(a) Lake made by removing seven resistors. 
		(b, c) Two different configurations when only five resistors are removed.
		The dangling bonds are represented by magenta lines.
		The resistance is calculated between sites $\mathbf{r}_i$ and $\mathbf{r}_j$ (blue dots).
		The red dot is the origin of the coordinate system.
		\label{bonds_2x3hole:fig}}
\end{figure}

In this section, as an example, we examine a topological defect where seven resistors are removed, resulting in two isolated sites and the creation of a lake, as shown in figure~\ref{bonds_2x3hole:fig}(a).
As discussed in section~\ref{gen_theory:sec}, due to the two isolated sites, the matrix $\mathbf{B}$ 
in equation~(\ref{G_ri_rj:eq}) becomes singular, and one should add dangling bonds in the lake.
Two possible configurations of dangling bonds passing through the two isolated sites are shown 
in figures~\ref{bonds_2x3hole:fig}(b), (c).
Since the current should be zero along these dangling bonds, they do not alter the resistance between the sites located outside the lake in the perturbed network. 
Moreover, both configurations can be used in the calculations without affecting the resistance.
To calculate the resistance, e.g., between sites 
$\mathbf{r}_i =\left(0,2\right)$ and $\mathbf{r}_j =\left(3,0\right)$, 
we now choose the configuration shown in figure~\ref{bonds_2x3hole:fig}(c). 
Five bonds are removed, consequently the dimension of the matrix $\mathbf{B}$ is five. 

After similar calculations as before, the effective resistance between 
sites $\mathbf{r}_i =\left(0,2\right)$ and 
$\mathbf{r}_j =\left(3,0\right)$  becomes (for details see~\ref{Hole_2_missing_nodes:app}): 
\begin{align}
	\label{R_res_fig3:eq}
	R(r_i,r_j) &= \frac{4}{3}+\frac{4}{9 (4-\pi )}-\frac{64}{3 \pi }
	+\frac{4}{3 (3 \pi-4)}+\frac{512}{9 \pi^2}
	\approx 1.0703...
\end{align}
It is worth comparing this value with the effective resistance in a perfect lattice between the same two sites:
$R_0 (3,2) = R_0 (2,3)=\frac{1}{2} + \frac{4}{3\pi} \approx  0.9244 ...$

\begin{figure}[hbt]
	\centering 
	\includegraphics[scale=0.4]{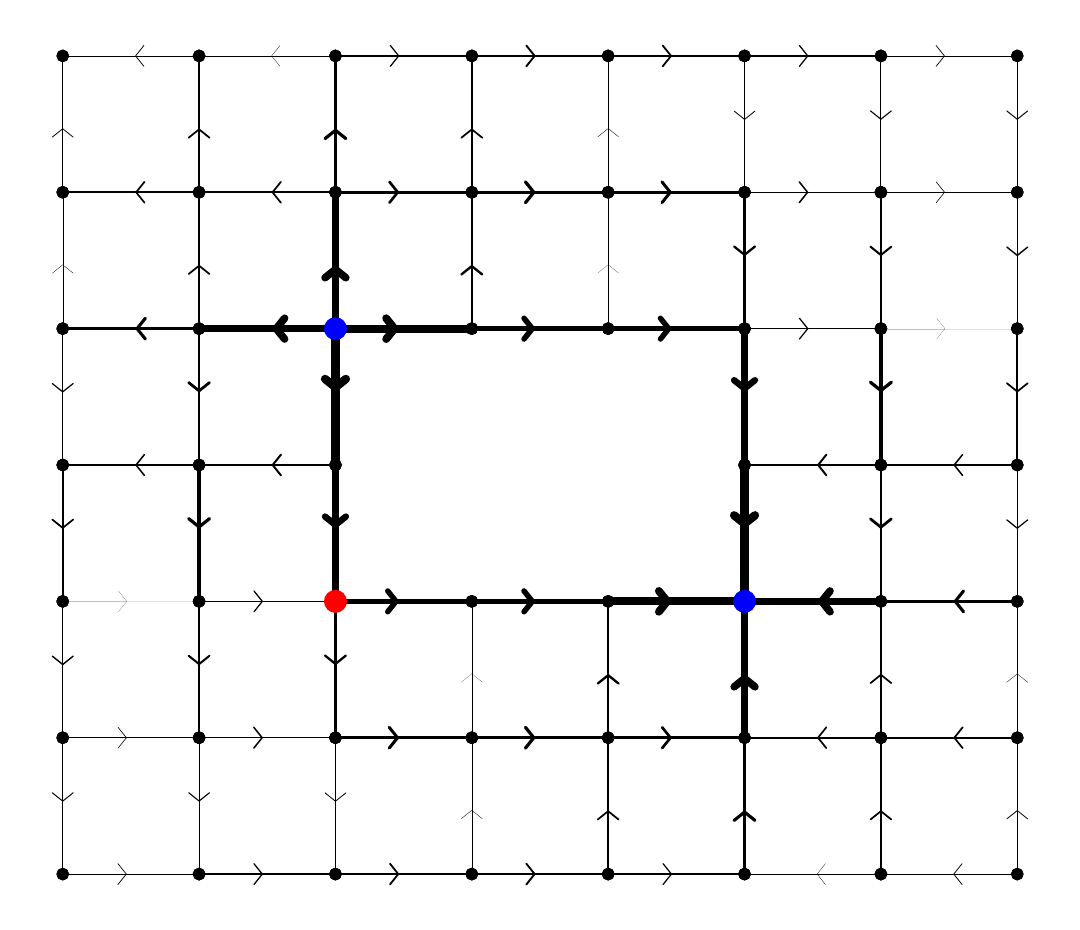}
	\caption{Bond current distribution in the perturbed lattice shown in figure~\ref{bonds_2x3hole:fig}(c). No current flows on the two dangling bonds inside the lake.
		\label{bonds_2x3hole_current:fig}}
\end{figure}

To see the bond current distribution, we calculated them using equation~(\ref{current:eq}) and are plotted in figure~\ref{bonds_2x3hole_current:fig}. 
As expected, the current is zero inside the lake.

\subsection{An island, as a topological defect}
\label{topo_islands:sec}

\begin{figure}[hbt]
	\centering 
	\includegraphics[scale=0.6]{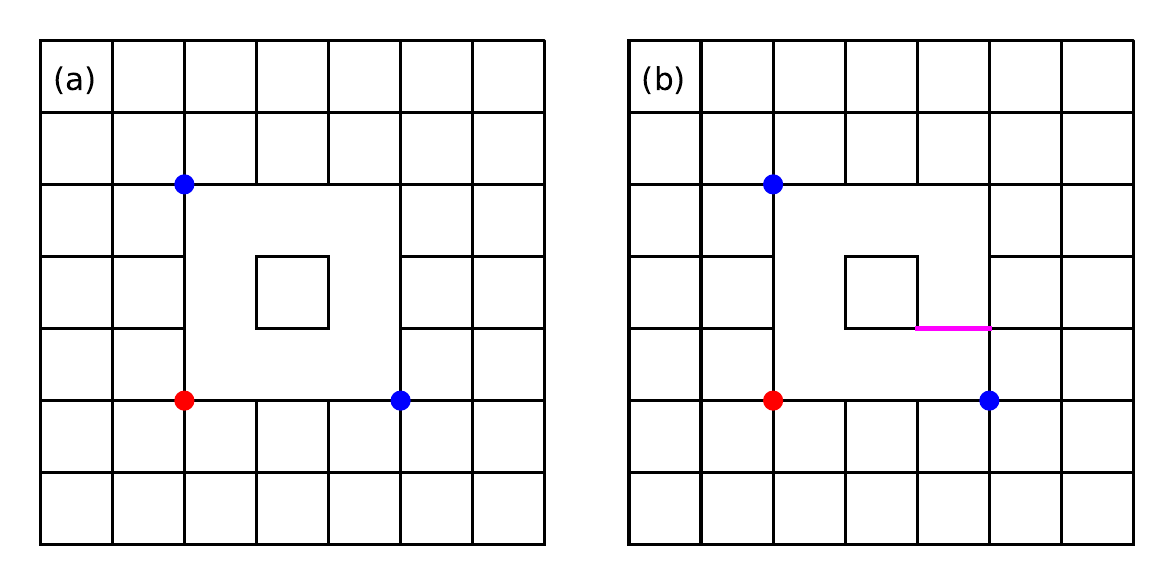}
	\caption{(a) Island created by removing 8 bonds. 
		(b) Bridge (magenta line) between the island and the rest of the lattice by removing only 7 bonds.
		The resistance is calculated between sites $\mathbf{r}_i$ and $\mathbf{r}_j$ (blue dots).
		The red dot is the origin of the coordinate system.
		\label{bonds_island_hole:fig}}
\end{figure}

In this section, we consider another type of topological defect, namely an island.
In our example, removing eight bonds, an island of resistors is created 
as shown in figure~\ref{bonds_island_hole:fig}(a).
As explained in section~\ref{gen_theory:sec}, the matrix $\mathbf{B}$ in equation~(\ref{G_ri_rj:eq}) becomes invertible if we introduced a bridge bond, 
that connects the island and the rest of the lattice 
(one example is shown in figure~\ref{bonds_island_hole:fig}(b)). 
If the lattice sites where the current enters and exits the lattice are outside the perturbed region, then adding a bridge bond does not change the effective resistance because the current is zero through the bond. 
Note that the position where one bridge bond is added does not affect the effective resistance 
(however,  more than one bridge bond would change the effective resistance since the current through them is no longer zero). 
Now, our theory can be applied and 
from equations~(\ref{G_Woodbury:eq}),~(\ref{G_pq:eq}) and~(\ref{gu_vg:eq}), the effective resistance between 
sites $\mathbf{r}_i =\left(0,3\right)$ and 
$\mathbf{r}_j =\left(3,0\right)$ is  given by 
\begin{align}
	R(r_i,r_j)  &= -36+\frac{961}{9 (8-\pi )}+\frac{2048}{45 \pi}+\frac{1}{3 \pi -8}
	\approx  1.1664...
\end{align}
For comparison, in a perfect square lattice, the resistance is 
$R(3,3) = \frac{46}{15 \pi} \approx 0.97615... $.

\begin{figure}[hbt]
	\centering 
	\includegraphics[scale=0.4]{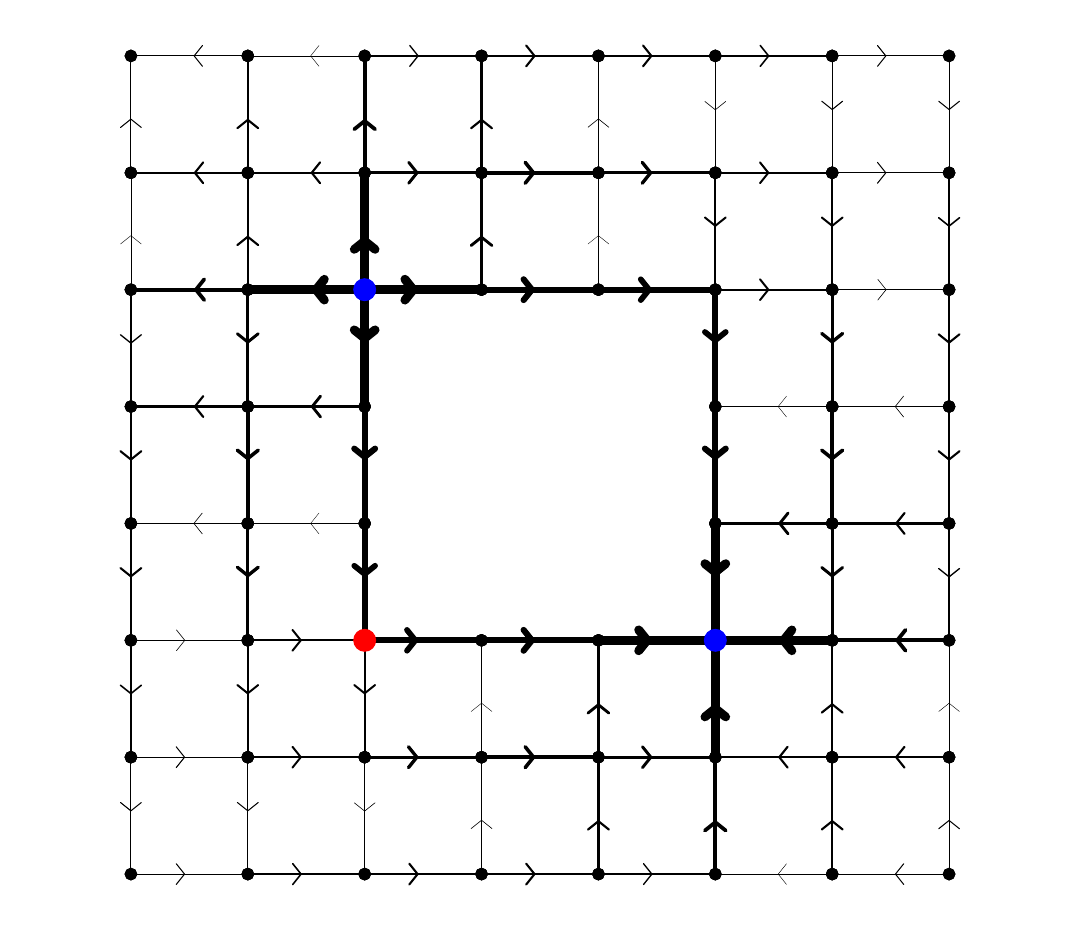}
	\caption{Bond current distribution in the perturbed lattice shown 
		in figure~\ref{bonds_island_hole:fig}(a). No current flows on the bonds inside the island.
		\label{bonds_island_current:fig}}
\end{figure}

Using equation~(\ref{current:eq}), we plotted the bond current distribution of the perturbed lattice in figure~\ref{bonds_island_current:fig}.
As expected, the current through the bridge to the island is zero.
Note that if there is no island, but only a lake with isolated sites, then the current distribution remains the same when the entry and exit sites for the current are located in the infinite part of the network. 

\subsection{Periodic defects}

\begin{figure}[hbt]
	\centering 
	\includegraphics[scale=0.4]{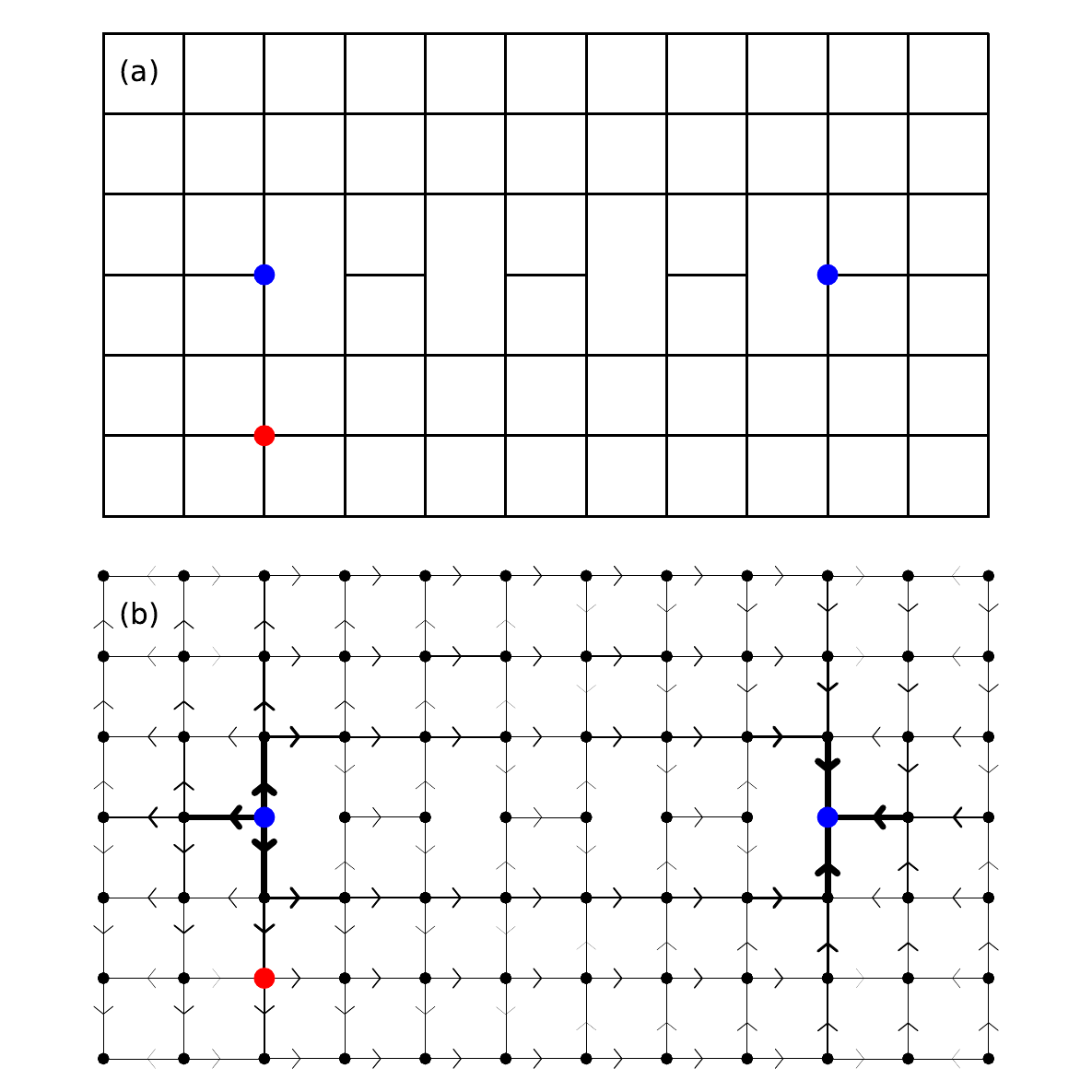}
	\caption{(a) Four bonds are removed periodically.
		(b)	Bond current distribution for the perturbed lattice. 
		\label{bonds_periodic_4_current:fig}}
\end{figure}

Figure~\ref{bonds_periodic_4_current:fig}(a) shows the perturbation of an infinite square lattice by removing four bonds periodically. 
Using our theory, one can find analytically that the effective resistance between sites  
$\mathbf{r}_i =\left(0,2\right)$ and 
$\mathbf{r}_j =\left(7,2\right)$ is 
\begin{align}
	R(r_i,r_j) &= \frac{9689625 \pi^3 -76600800 \pi^2 +172254624 \pi -85590016 }
	{30 \pi  \left(37665 \pi^2-96528 \pi -68608\right)}
	\approx  1.4515 ... 
\end{align}
It is worth comparing this value with the effective resistance in a perfect lattice between the same two sites: 
$R_0 (7,0) = \frac{11073}{2}-\frac{260848}{15 \pi } 
\approx  1.1335 ...$
For the perturbed lattice, the bond current distribution 
obtained from equation~(\ref{current:eq}) is  plotted in figure~\ref{bonds_periodic_4_current:fig}(b).

\begin{figure}[hbt]
	\centering 
	\includegraphics[scale=0.55]{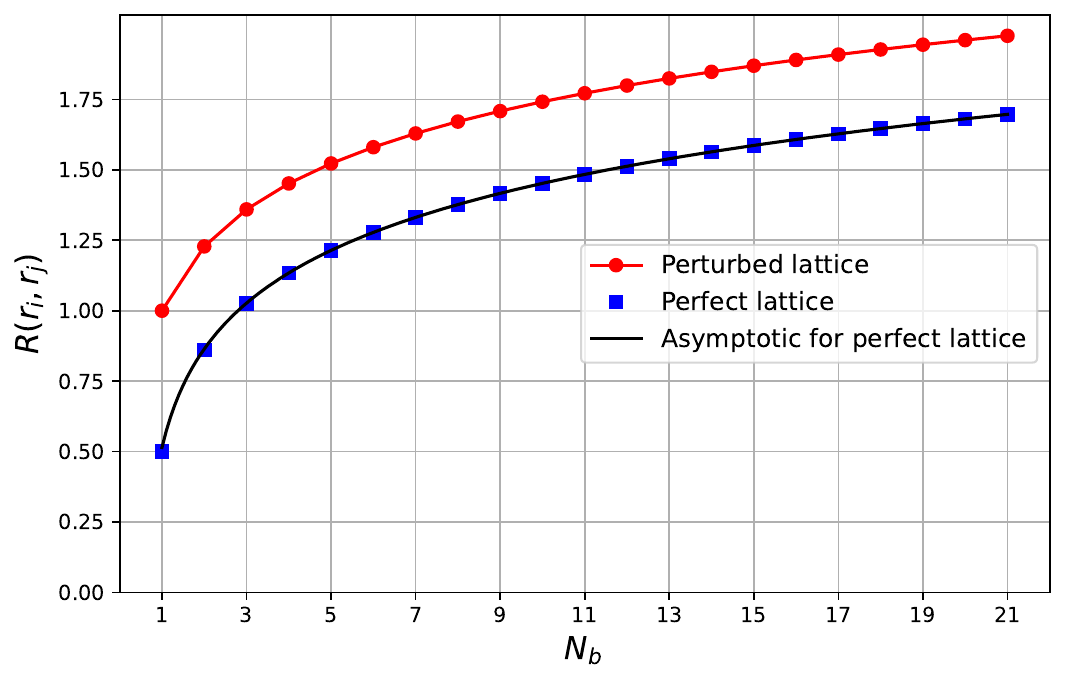}
	\caption{The resistance of the periodically perturbed square lattice 
		(red dots, while the red line are just for guiding the eyes), 
		and that for the perfect lattice (blue square symbols) 
		as a function of the number of removed resistors $N_b$. 
		The asymptotic values of the resistance for a perfect lattice, as calculated 
		from equation~(\ref{R_sq-asym:eq}) are also plotted (black solid line). 
		\label{bonds_periodic_R_distance:fig}}
\end{figure}

Figure~\ref{bonds_periodic_R_distance:fig} shows the effective resistance $R(r_i,r_j)$ as a function of the number of removed resistors in the periodically perturbed chain.
The current enters and exits at the sites at the two ends of the chain, that is,
at $\mathbf{r}_i =\left(0,2\right)$ and
$\mathbf{r}_j =\left(2 N_b-1,2\right)$, respectively, where
$N_b$ is the number of resistors removed from the chain. 
Figure~\ref{bonds_periodic_4_current:fig} shows the case $N_b=4$.
The asymptotic form of the resistance for a perfect square lattice is
\begin{align}
	R_0^\mathrm{asym}(m,n) &= \frac{R}{\pi} \left( \ln \sqrt{m^2+n^2} +\gamma + \frac{\ln	8}{2} \right),
	\label{R_sq-asym:eq}
\end{align}
where $\gamma = 0.5772\dots$ is the Euler-Mascheroni
constant\cite{Arfken:book}. 
Here the indices $m$ and $n$ are given by equation~(\ref{mn_indices:def}). 
This expression was derived in reference~\cite{sajat_perfect_AJP_10.1119/1.1285881}. 
We should emphasize that the agreement between the exact and asymptotic resistance values for the square lattice is remarkable, even for small separation of the sites where the current enters and exits from the lattice. 

\subsection{Perturbation of a triangular lattice}
\label{triangular:sec}

\begin{figure}[hbt]
	\centering 
	\includegraphics[scale=0.5]{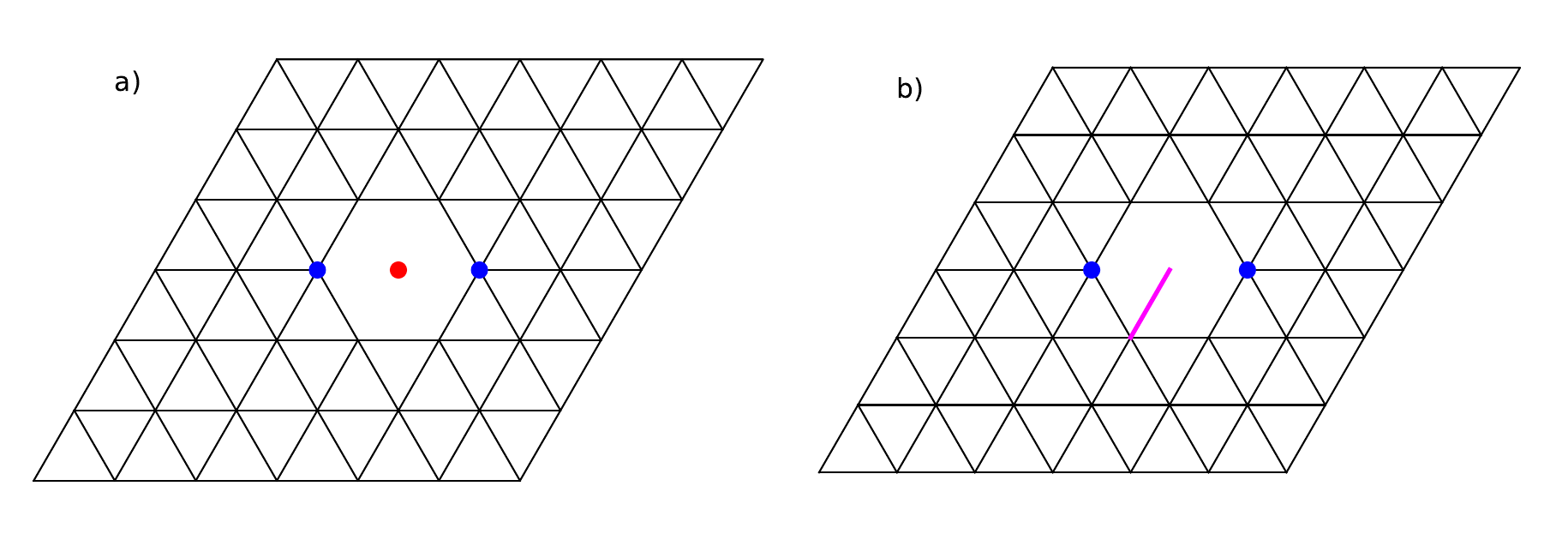}
	\caption{(a) Six bonds are removed, creating a hole in a triangular lattice. 
		(b) Dangling bond (magenta line) inside the hexagon, thus only five bonds are removed. The origin of the coordinate system is at the middle of the hexagon. 
		The resistance is calculated between sites $\mathbf{r}_i$  
		and $\mathbf{r}_j$  (blue dots).
		\label{6szog_missing_A:fig}}
\end{figure}

Our general theory can be extended to other perturbed lattices provided the effective resistances of the corresponding perfect lattice are known. Here, we focus on the perturbation of a perfect infinite triangular lattice.
As an example, we now consider the resistor network shown in figure~\ref{6szog_missing_A:fig}(a). 
In this perturbation, six bonds are removed, creating a hexagon-shaped hole, and thus, one site becomes isolated in the perfect triangular lattice.  
Such a topological defect, namely a lake, as discussed in section~\ref{gen_theory:sec},  
can be treated by introducing a single dangling bond as shown in figure~\ref{6szog_missing_A:fig}(b).
Then, only five bonds are removed, and the resistance does not change since no current flows through the remaining dangling bond. 
Moreover, the matrix $\mathbf{B}$ is no longer singular. 
Thus, our theory can be applied. 
The details of the calculations are in~\ref{3szog_hexagon:app}.
We find that the effective resistance between sites  
$\mathbf{r}_i$ and 
$\mathbf{r}_j$ shown in figure~\ref{6szog_missing_A:fig}(a) is 
\begin{align}
	\label{3szog_hexagon_missing_R:eq}
	R(r_i,r_j) &= 
	\frac{139968 \sqrt{3}-338256 \pi +90720 \sqrt{3} 
		\pi^2-23112 \pi^3-660 \sqrt{3} \pi^4+275 \pi^5}
	{2
		\left(-34992 \sqrt{3}+58320 \pi -7128 \sqrt{3} \pi^2-1512 
		\pi^3+165 \sqrt{3} \pi^4+25 \pi^5\right)}
	\approx  0.62156 ...
\end{align}
For comparison, in a perfect triangular lattice, the effective resistance 
is  $R_\triangle (r_i,r_j) = 
\frac{8}{3}-\frac{4 \sqrt{3}}{\pi } \approx  0.461351...  $ 

Note that knowing the matrix $\mathbf{B}$, one can easily calculate the resistance between other lattice points, too, only the vectors $\mathbf{U}^{\left(i\right)}$ and $\mathbf{V}^{\left(j\right)}$ need to be recalculated. 

It is known that, similarly to the square lattice, the two‑point resistance on a perfect triangular lattice grows logarithmically with node separation~\cite{Azimi-Tafreshi_2010_saym_triangular,Owaidat_2018_asym_triangular}; hence we expect the perfect and periodically defective (perturbed) lattices to show analogous asymptotic behaviour to that in figure~\ref{bonds_periodic_R_distance:fig} for the square lattice.

\section{Numerical results  }
\label{num_examples:sec}

Our analytical calculations of resistances can be readily implemented as an algorithm.
In this section, we provide examples to illustrate the calculation of effective resistance in more complex perturbed resistor networks.

\subsection{Obstacle}

\begin{figure}[hbt]
	\centering 
	\includegraphics[scale=0.5]{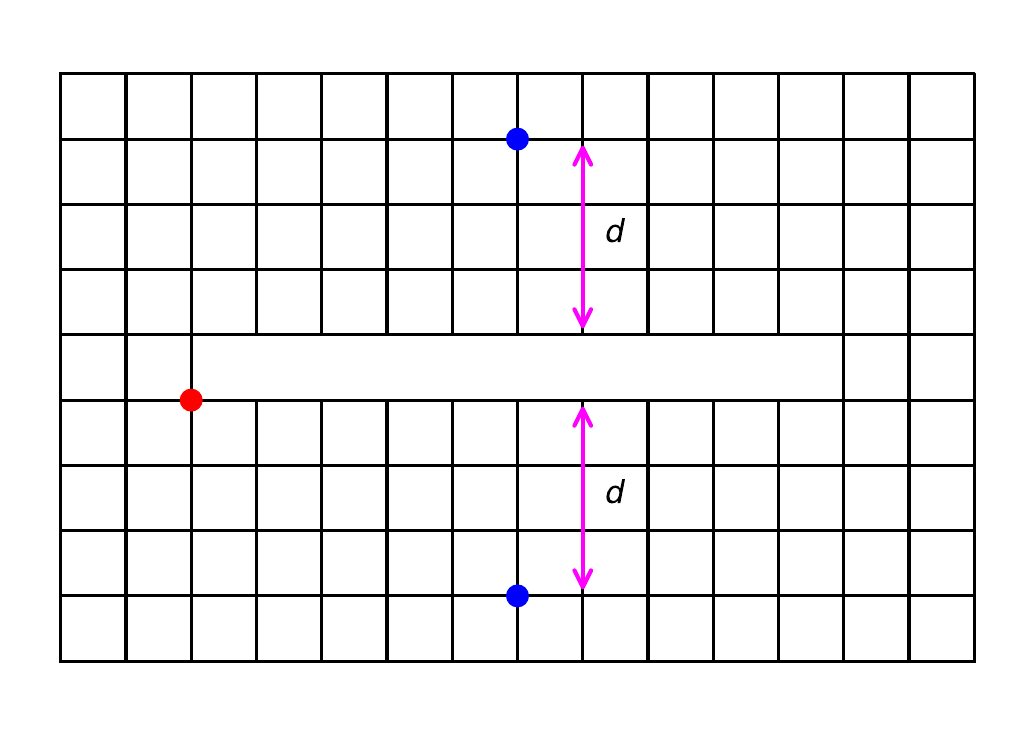}
	\caption{
		An obstacle in the square lattice is created by removing 9 resistors. 
		The resistance is calculated between sites $\mathbf{r}_i=(5,-d)$ and 
		$\mathbf{r}_j=(5,d+1)$ (blue dots),  
		which are located at a lattice distance $d$ from the obstacle.
		In this figure $d=3$.
		The red dot is the origin of the coordinate system.
		\label{bonds_obstacle_setup_arrows:fig}}
\end{figure}

Figure~\ref{bonds_obstacle_setup_arrows:fig} shows the perturbation of an infinite square lattice by removing 9 bonds. 
The two-point resistance is calculated between $\mathbf{r}_i$ and $\mathbf{r}_j$, 
which lie on the vertical symmetry axis of the obstacle, and their lattice distance from the obstacle is $d$, where $\mathbf{r}_i=(5,-d)$ and $\mathbf{r}_j=(5,d+1)$, and $d\ge 0$ is an integer. 
From our numerical calculations, we found that $R(d=0) =1.9803...$ 

\begin{figure}[hbt]
	\centering 
	\includegraphics[scale=0.55]{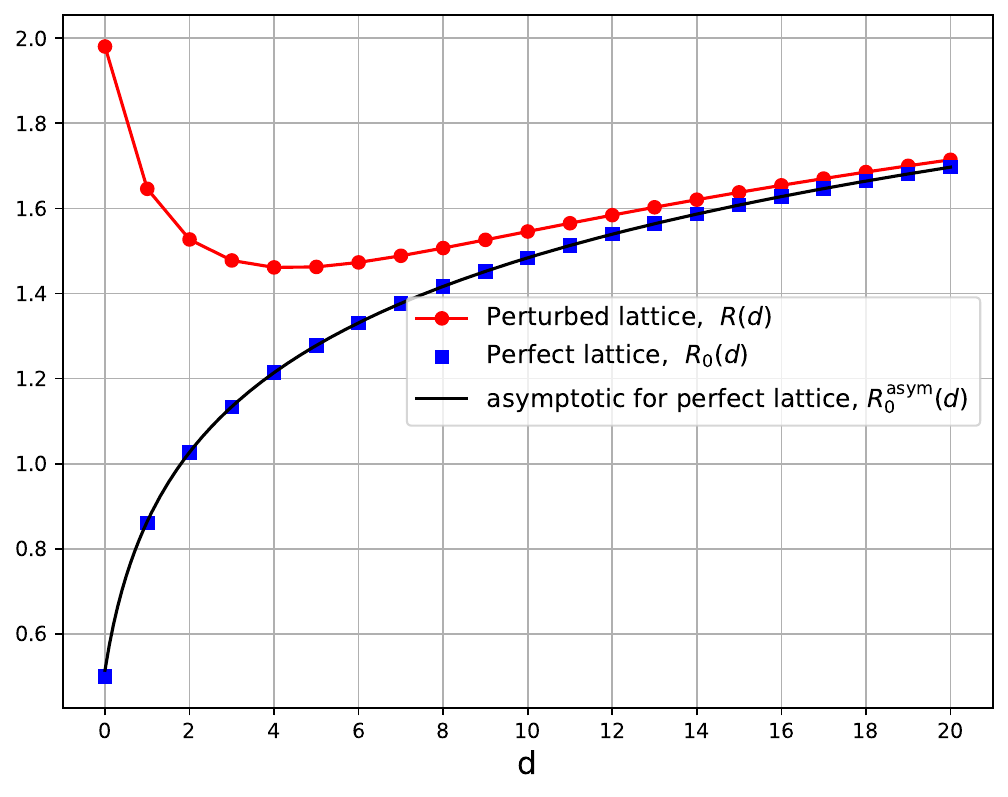}
	\caption{
		The two-point resistance $R(d)$ for an obstacle shown in figure~\ref{bonds_obstacle_setup_arrows:fig}  
		as a function of the distance $d$ 
		(red dots, while the red line are just for guiding the eyes), 
		and that for the perfect lattice (blue square symbols). 
		The asymptotic values of the resistance for a perfect lattice, as calculated 
		from equation~(\ref{R_sq-asym:eq}) are also plotted (black solid line).
		\label{obstacle_R_d:fig}}
\end{figure}

Figure~\ref{obstacle_R_d:fig} shows the effective resistance $R(d)\equiv R(r_i,r_j)$ as a function of the distance $d$.
One can see that the resistance $R(d)$ exhibits an unusual dependence. 
Generally, in a square lattice, the resistance increases with the separation between the two points $\mathbf{r}_i$ and $\mathbf{r}_j$. 
However, when an obstacle is introduced into the perfect lattice, one can see from the figure that the resistance initially decreases, reaching a minimum ($d = 4$ in the figure), and then increases 
for increasing $d$, ultimately approaching the asymptotic value characteristic of the perfect lattice. 
This behavior of $R(d)$ can be understood qualitatively from equation~(\ref{R_ij_perturbed:eq}), where the two-point resistance is expressed as the sum of two terms: the first, $R_0(r_i,r_j)$, the resistance in the perfect lattice, while the second term accounts for the excess resistance resulting from the removal of bonds. 
This second contribution is less relevant when the two points, $\mathbf{r}_i$ and $\mathbf{r}_j$, 
move farther from the obstacle, resulting in a monotonically decreasing function of $d$.
From equation~(\ref{R_sq-asym:eq}) it is clear that $R_0(r_i,r_j)$ is monotonically increasing function of $d$.
Consequently, the sum of these two terms in equation~(\ref{R_ij_perturbed:eq}) leads to a minimum in the resistance as a function of $d$. 
Note that when the length of the obstacle is increased (in the figure it is 9) then the minimum position of the resistance becomes larger.

\subsection{A picture of a car made of wires}
\label{car:sec}

\begin{figure}[hbt]
	\centering 
	\includegraphics[scale=0.5]{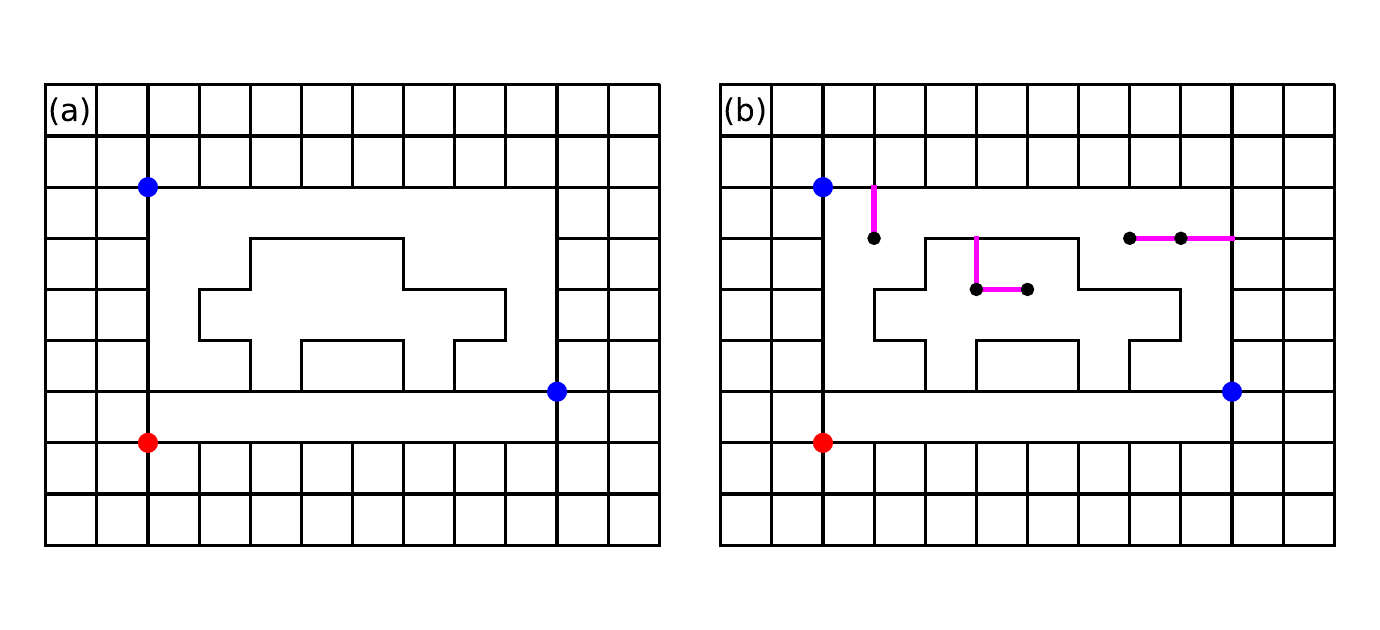}
	\caption{(a) A `car' made of wires. The number of removed resistors is 41. 
		(b) The network is connected by introducing dangling bonds (magenta lines).
		The resistance is calculated between sites $\mathbf{r}_i$ and $\mathbf{r}_j$ (blue dots).
		The red dot is the origin of the coordinate system.
		\label{car_sq:fig}}
\end{figure}

Resistors removed from an infinite square lattice form the shape of a car, 
as shown in figure~\ref{car_sq:fig}(a).
As discussed in section~\ref{gen_theory:sec}, there are isolated nodes in the perturbed lattice, 
and so we need to add dangling bonds. One possibility is depicted in figure~\ref{car_sq:fig}(b).

In this case, 41 bonds are removed, therefore, the matrix $\mathbf{B}$ of size 41 should be inverted.
From our numerical calculation, we find that the effective resistance between 
sites $\mathbf{r}_i =\left(8,1\right)$ and 
$\mathbf{r}_j =\left(0,5\right)$  is $R(r_i,r_j) \approx 1.5130...$
For comparison, in a perfect square lattice, the effective resistance 
is $R_0(r_i,r_j) \approx 1.2122...$

\subsection{`JPA 2025' characters}
\label{JPA_2025:sec}

\begin{figure}[hbt]
	\centering 
	\includegraphics[scale=0.5]{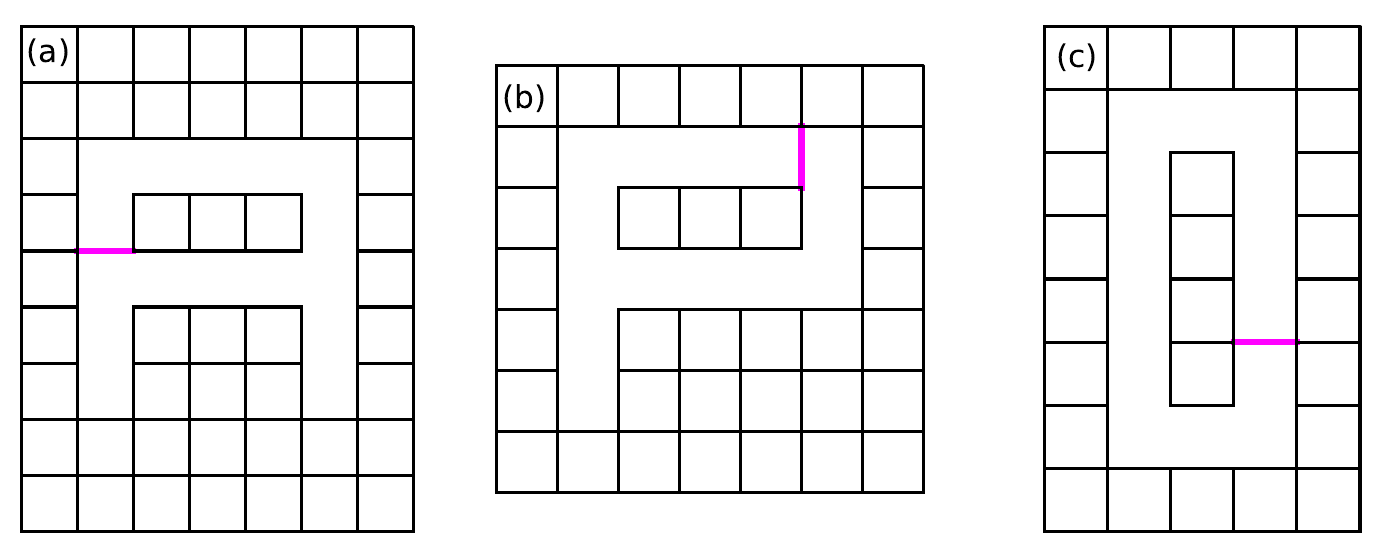}
	\caption{Alphanumeric characters \textit{A}, \textit{P} and \textit{0} contain an island of resistors (Figures~(a), (b), (c), respectively).
		In the modified network, the island in each character is connected to the rest of the lattice with a bridge bond (magenta lines).
		\label{A_P_0_mod:fig}}
\end{figure}

We now consider the resistance of the perturbed resistor network between the two grid points shown in figure~\ref{JPA:fig} in the introduction. 
However, one should be careful when treating the alphanumeric characters \textit{A,P} and \textit{0} since each of them contains islands.
As discussed in section~\ref{gen_theory:sec}, one should add a bridge bond between each island and the rest of the network.
One possibility is plotted for each character in figure~\ref{A_P_0_mod:fig}. 
Thus, in figure~\ref{JPA:fig} the number of removed bonds is 81.
After such network modification, one can calculate the resistance between 
any two lattice sites of the perturbed lattice. 
In this way, we find the resistance given in the introduction, i.e., $R(r_i,r_j) = 1.6645...$. 

\section{Conclusions}
\label{summary:sec}

Since the perturbation of the resistor network is related to a sum of dyadic products of the unperturbed lattice,  we can apply the Woodbury matrix identity to obtain the Green operator of the perturbed network.  
We demonstrated that this approach provides a much simpler and more efficient treatment of this problem than the recursive application of the change to the bonds one by one.
Indeed, by removing (or replacing) more than one bond in the resistor network, all bonds involved in the perturbation can be treated simultaneously.
We presented numerous examples of resistor networks to obtain the two-point resistance and the bond current distribution in a perturbed infinite square and triangular lattice.  
We showed that our formalism can be easily used analytically when only a few bonds are removed.  
Based on our experience, the recursive algorithm is more involved to implement and tends to accumulate larger rounding errors through repeated updates. 
In contrast, our method is easier to code and more numerically stable.

Our method can be used with other types of resistor networks, such as honeycomb lattices, cubic lattices, and more complex tilings with resistors studied, e.g., in reference~\cite{sajat_tiling_2011}, 
as well as the half- and quarter regions of the square resistor lattice~\cite{half_quater_image_1_8821548,half_quater_image_2_9078361}.
The key requirement for these generalizations is the knowledge of Green operator $\hat{G}_0$ for the unperturbed lattice.

We can study not only infinite networks but also perturbations of finite resistor networks with our approach. 
Moreover, our method can be useful for locating missing resistors in a finite resistor network by measuring the resistance on the boundary of the system. This issue is rather important in the failure of a power system. 
Furthermore, our formalism can be extended to $RLC$ circuits, providing a framework for studying the topological insulating and semimetallic states realized in topoelectric circuits~\cite{Topo_Molenkamp:cikk,Topo_PhysRevB.107.245114,topo_APL_Electronic_1063/5.0265293_2025}. 

Finally, we emphasize the importance of using proper mathematical notation to enhance the transparency and insight of the formalism. 
The introduction of the bra-ket notation, inspired by quantum mechanics, makes the description of our concepts more intuitive and accessible.
In summary, our work provides a robust and versatile framework for studying lattice perturbations not only in the problem of resistor networks but in other research fields of physics, engineering, and materials science.

\ack

I would like to thank Anthony J.~Guttmann, Colin J.~Lambert, 
Merse Előd Gáspár, Róbert Németh and Gábor Széchenyi for their helpful advice during the preparation of the manuscript.
This project is supported by the National Research, Development, and Innovation Office
(NKFIH, Grant No. K134437), and the TRILMAX Horizon Europe consortium (Grant No. 101159646).

\appendix

\section{Derivation of the Green operator of the perturbed system}
\label{Green:app}

In this section, using the Woodbury identity (\ref{Woodbury:eq}) 
we derive equation~(\ref{G_Woodbury:eq}), i.e., 
the inverse in equation~(\ref{G_a:eq}).
The $i,j$ matrix element of operator $\hat{L}_1$ is given by  
\begin{align}
	\label{L_ij_app:eq}
	(L_1)_{ij} &= \langle i| \hat{L}_1 |j \rangle 
	= \sum_{p=1}^N \langle i | b_p \rangle g_p \langle b_p | j \rangle
	=\sum_{p,q=1}^N \langle i | b_p \rangle  
	g_p \delta_{pq} \langle b_q| j \rangle \nonumber \\[2ex]
	&= \sum_{p,q=1}^N  U_{ip} C_{pq} V_{qj} 
	= {\left(\mathbf{U} \mathbf{C}\mathbf{V}\right)}_{ij} , 
\end{align}
where 
$U_{ip} = \langle i | b_p \rangle $,   
$V_{qj} = \langle b_q| j \rangle $,  
$C_{pq} = g_p \delta_{pq}$.
Note that here indices takes the values $p, q =1, 2, \dots, N$, 
and $i$ and $j$ run over all the lattice points (finite or infinite).
Thus, $\mathbf{C}$ is a $N \times N$ diagonal matrix, 
while $\mathbf{U}$ and $\mathbf{V}$ are rectangular matrices 
with $N$ columns and rows, respectively. 
The matrix representation of the operator $\hat{L}_1$ is 
$\mathbf{L}_1 = \mathbf{U} \mathbf{C}\mathbf{V}$.   

Now from the Woodbury identity (\ref{Woodbury:eq})
we find that the matrix representation of $\hat{G}$ is 
\begin{subequations}
	\begin{align}
		\mathbf{G} &= -{\left( \mathbf{L}_0 + \mathbf{L}_1 \right)}^{-1} 
		= -{\left( \mathbf{L}_0 + \mathbf{U} \mathbf{C}\mathbf{V} \right)}^{-1}
		= - \mathbf{L}_0^{-1} 
		+ \mathbf{L}_0^{-1}\mathbf{U} \mathbf{B}^{-1} 
		\mathbf{V} \mathbf{L}_0^{-1}  \nonumber \\[2ex]
		&= \mathbf{G}_0 
		+ \mathbf{G}_0  \mathbf{U} \mathbf{B}^{-1} \mathbf{V} \mathbf{G}_0 ,
		\quad \textrm{where}  \\[2ex] 
		\mathbf{B} &= \mathbf{C}^{-1}
		\label{B_def:eq}
		+ \mathbf{V}  \mathbf{L}_0^{-1}\mathbf{U} 
		= \mathbf{C}^{-1}
		- \mathbf{V}  \mathbf{G}_0\mathbf{U} .
	\end{align}
\end{subequations}
To make the expressions $\mathbf{G}$ and $\mathbf{B}$ 
more transparent, it is helpful to use the bra-ket formalism.  
For fixed indices $i$ and $j$ the matrix element of $\mathbf{G}$ 
can be written as   
\begin{subequations}
	\begin{align}
		G_{ij} 	&= {\left(G_0\right)}_{ij} 
		+ \sum_{p,q=1 }^N \sum_{ k,t} {\left(G_0\right)}_{ik}
		U_{kp} {\left(\mathbf{B}^{-1}\right)}_{pq} V_{qt} {\left(G_0\right)}_{tj} 
		\nonumber \\[2ex]
		&= \langle i| \hat{G}_0 |j \rangle 
		+ \sum_{p,q=1 }^N \sum_{ k,t}
		\langle i| \hat{G}_0 |k \rangle  
		\langle k| b_p \rangle 
		{\left(\mathbf{B}^{-1}\right)}_{pq}
		\langle b_q| t \rangle
		\langle t| \hat{G}_0 |j \rangle  \nonumber  \\[2ex]
		&= 
		\langle i| \hat{G}_0 |j \rangle 
		+ \sum_{p,q=1 }^N 
		\langle i| \hat{G}_0 | b_p \rangle 
		{\left(\mathbf{B}^{-1}\right)}_{pq}
		\langle b_q| \hat{G}_0 | j \rangle  \nonumber   \\[2ex]
		&= 
		\label{Gij_A_app:eq}
		\langle i| \hat{G}_0 | j \rangle 
		+ 
		\langle i| \hat{G}_0 | \alpha \rangle \,
		\mathbf{B}^{-1} \,
		\langle \alpha| \hat{G}_0 | j \rangle     \\[2ex]
		&= 
		\label{Gij_B_app:eq}
		\langle i| \hat{G}_0 |j \rangle 
		+ \mathbf{U}^{\left(i\right)} \,  
		\mathbf{B}^{-1}  \, 
		\mathbf{V}^{\left(j\right)} ,  
	\end{align}
\end{subequations}
where the summation $k$ and $t$ are over the entire lattice points 
(for finite or infinite lattice), 
and thus, the summation over these indices can be performed using the completeness relation 
$\sum_k |k \rangle \langle k| = \hat{\mathbb{1}}$, 
where $\hat{\mathbb{1}}$ is the identity operator. 
In equation~(\ref{Gij_A_app:eq}) we used the definition of the ket vector $|\alpha \rangle$ given 
in equation~(\ref{alpha:def}) in the main text. 
In the last step, we introduce two vectors $\mathbf{U}^{\left(i\right)}$ and $\mathbf{V}^{\left(j\right)}$,  which are $N$-component row and column vectors, respectively, as defined in 
equations~(\ref{ri_G0_a:eq}) and~(\ref{a_G0_rj:eq}) in the main text.  
Note that the vectors $\mathbf{U}^{\left(i\right)} $ and $\mathbf{V}^{\left(j\right)} $ are different from the matrices $\mathbf{U}$ and $\mathbf{V}$ in Eq~(\ref{L_ij_app:eq}). 
The previous two depend on $\mathbf{r}_i$ and $\mathbf{r}_j$,  respectively.

We still need to calculate the matrix elements of the matrix $\mathbf{B}$. 
For $m, n = 1,2,\dots , N$ from equation~(\ref{B_def:eq}) we have 
\begin{subequations}
	\begin{align}
		B_{mn} &= g^{-1}_m \delta_{mn} 
		-\sum_{s,t} V_{ms} {\left(G_0\right)}_{st} U_{tn} 
		= g^{-1}_m \delta_{mn} 
		- \sum_{s,t}  \langle b_m| s \rangle
		\langle s| \hat{G}_0 | t \rangle \langle t | b_n \rangle 
		\nonumber \\[2ex]
		&=   g^{-1}_m \delta_{mn} - \langle b_m| \hat{G}_0 | b_n \rangle , 
	\end{align}
	where, in the last step, we used the completeness relation again. 
	Using the ket vector $|\alpha\rangle$ it is clear that $B_{mn}$ is the matrix element of a finite, $N\times N$ matrix 
	\begin{align}
		\label{B_matrix_app:eq}
		\mathbf{B} &=  
		\mathbf{C}^{-1} - \langle \alpha |\,\hat{G}_0\, |\alpha \rangle .
	\end{align}
\end{subequations}

Finally,  equation~(\ref{Gij_B_app:eq}), together with equation~(\ref{B_matrix_app:eq}) 
are the ones provided in the main text, in equation~(\ref{G_Woodbury:eq}).
Moreover, from equation~(\ref{Gij_A_app:eq}) it can be seen that the operator $\hat{G}$ is the same as the one given in the main text, in equation~(\ref{G_operator:eq}).

\section{Relation between the energy-dependent Green operator in quantum mechanics and that in resistor networks}
\label{G_quantum:app}

In this section, we present an alternative derivation of our main result, given in equation~(\ref{G_operator:eq}). This approach avoids the use of the Woodbury identity and may be more accessible to physicists familiar with perturbation theory based on Green's functions.

First, note that the Green operator $\hat{G}=-\hat{L}^{-1}$ 
associated with the Laplacian operator, as defined in equation~(\ref{Green:eq}), is related to the energy-dependent Green operator used in quantum 
mechanics~\cite {economou2006green}:
\begin{align}
	\hat{G}(E) &= {\left( E\,\hat{\mathbb{1}}-\hat{H}\right)}^{-1}
	\label{G_QM:def}
\end{align}
by setting $E=0$ and replacing $\hat{H}$ by $\hat{L}$.
This analogy allows us to apply the standard perturbation theory commonly used in quantum mechanics. In fact, the derivation presented below can be directly utilized in quantum mechanical contexts.

The one-particle Hamiltonian $\hat{H}= \hat{H}_0 + \hat{H}_1$ can be separated into an unperturbed part $\hat{H}_0$ and a perturbation $\hat{H}_1$.
Using equation~(\ref{G_QM:def}) one can write that 
$\left(\hat{G}_0^{-1} - \hat{H}_1\right)\hat{G}  =\hat{\mathbb{1}}$, 
where $\hat{G}_0(E)  = {\left( E\,\hat{\mathbb{1}}-\hat{H}_0\right)}^{-1}$.
Then multiplying this equation on the left by $\hat{G}_0$ and rearranging yields 
the well-known Dyson equation~\cite{economou2006green}:
\begin{align}
	\hat{G} &= \hat{G}_0 + \hat{G}_0 \hat{H}_1 \hat{G}.
	\label{Dyson:def}
\end{align}  
For simplicity, we suppress the explicit $E$-dependence of $\hat{G}$ and $\hat{G}_0$. 
This equation can be expanded into an infinite series by repeatedly substituting the equation into itself:
\begin{align}
	\hat{G}(E) &= \hat{G}_0 + \hat{G}_0 \hat{H}_1 \hat{G}_0 +
	\hat{G}_0 \hat{H}_1 \hat{G}_0 \hat{H}_1 \hat{G}_0 + \dots . 
	\label{G_series:def}
\end{align}  
Since $\hat{H}_1$ has the special dyadic form, this infinite series can be summed, so the Dyson equation~(\ref{Dyson:def}) can be solved in closed form. 
Indeed, inserting equation~(\ref{G_a:eq}), i.e., $\hat{H}_1 =|\alpha \rangle \mathbf{C} \langle \alpha |$ into  equation~(\ref{Dyson:def}) yields
\begin{align}
	\hat{G} &= \hat{G}_0 
	+ \hat{G}_0 |\alpha \rangle \mathbf{C} \langle \alpha | \hat{G}.
	\label{Dyson_2:def}
\end{align}  
Now multiplying this equation from the left by $\mathbf{C} \langle \alpha |$ and rearranging the equation, we find
\begin{align}
	\mathbf{C} \langle \alpha | \hat{G} &= 
	{\left[
		\mathbf{C}^{-1}- \langle \alpha | \hat{G}_0 |\alpha \rangle 
		\right]}^{-1} \langle \alpha | \hat{G}_0.
	\label{Dyson_3:def}
\end{align}  
Note that 
$ {\left[\mathbf{C}^{-1} 
	- \langle \alpha | \hat{G}_0 |\alpha \rangle \right]}^{-1}$ 
is a finite-dimensional matrix whose dimension equals the number of removed bonds. 
Then, inserting equation~(\ref{Dyson_3:def}) 
into~(\ref{Dyson_2:def}) we obtain the final result:
\begin{align}
	\hat{G}(E) &= \hat{G}_0 + \hat{G}_0 |\alpha \rangle
	{\left[
		\mathbf{C}^{-1}- \langle \alpha | \hat{G}_0 |\alpha \rangle 
		\right]}^{-1} 
	\langle \alpha | \hat{G}_0 .
	\label{G_final:def}
\end{align}  
By setting $E=0$ and replacing $\hat{H}$ with $\hat{L}$, i.e. 
$\hat{G}=-\hat{L}^{-1}$ (and likewise $\hat{G}_0=-\hat{L}_0^{-1}$), 
one obtains the result given by equation~(\ref{G_operator:eq}).

Note that equation~(\ref{G_final:def}) can also be used to treat disordered systems containing impurities~\cite{economou2006green}. 
Moreover, our approach has been applied to a cluster of impurities in dynamical mean-field theory~\cite{RevModPhys.68.13,Senechal2012,PhysRevLett.87.186401,PhysRevB.62.R9283}, 
and we expect it can be used to obtain analytic results for the critical behavior of the planar Ising model with defects~\cite{PhysRevE.92.032108,PhysRevE.95.052101}.

\section{Green operator for finite network}
\label{G_finite:app}

For a finite network, Wu derived an expression for the two-point resistance in terms of the eigenvectors of the Laplacian matrix $\mathbf{L}$~\cite{Wu_R_eigenv_2004}.
Based on the concept of resistance distance introduced by
Klein and Radi\'{c}~\cite{Klein_Radic_1993} and further explored in references~\cite{Babic_Klien_application_2002,Xio_Gutman_2003,Bapat_resistance_wighted_2004}, 
we proposed an alternative method that eliminates the need to compute the Laplacian’s eigenvectors. 

The problem is that in equation~(\ref{Rij:eq}) the Green operator cannot be used since there is no inverse of the Laplacian matrix, neither for a perfect nor a perturbed finite network.
To see this, consider the relation between the vectors $\mathbf{I}$ and $\mathbf{V}$ 
formed from the currents and potentials at the lattice sites. 
Then, from the Kirchhoff’s and Ohm's laws it follows that $\mathbf{L}\,\mathbf{V} = -\mathbf{I}$, 
where $\mathbf{L}$ is the $N \times N$ Laplacian matrix of a finite network with $N$ nodes 
(for simplicity, we assume that the equation is already in dimensionless form). 
Define the all-ones $N$‑dimensional vector 
$\mathbf{f} ={\left(1,1,\dots,1\right)}^T$.
The current conservation implies that the sum of all rows (or columns) of $\mathbf{L}$ is identically zero, consequently one eigenvalue of the Laplacian is zero with eigenvector $\mathbf{f}$, 
so it is singular; its inverse, related to the Green operator, does not exist.

However, if a constant value is added to the voltage at each node in the finite network, the current distribution does not change. This fact allows us to determine the Green operator used for calculating the two-point resistance or impedance in a finite connected network.
To this end,  introduce the modified Laplacian
$\mathbf{L}^{\prime} = \mathbf{L} + \mathbf{f}\circ \mathbf{f}$.
We claim that the solution $\mathbf{V}^\prime $ of equation $\mathbf{L}^{\prime}\, \mathbf{V}^\prime = -\mathbf{I}$ differs from the original potential vector $\mathbf{V}$ only by a constant shift at every node, namely 
\begin{align}
	\mathbf{V}^\prime  &= 
	\mathbf{V} - \mathbf{f}\,\frac{\mathbf{f}\,\mathbf{V}}{\mathbf{f}^2}.
\end{align}
Indeed, using $\mathbf{L}\, \mathbf{f} = \mathbf{0}$ one finds
\begin{align}
	\mathbf{L}^{\prime}\, \mathbf{V}^\prime &= 
	\left(\mathbf{L} + \mathbf{f}\circ \mathbf{f} \right)
	\left(\mathbf{V}- \mathbf{f}\, \frac{\mathbf{f}\, \mathbf{V}}{\mathbf{f}^2} \right) 
	= \mathbf{L}\, \mathbf{V} = -\mathbf{I}.
\end{align}
Hence, the two  equations 
$\mathbf{L}\, \mathbf{V}= -\mathbf{I}$ and 
$\mathbf{L}^{\prime}\, \mathbf{V}^\prime = -\mathbf{I}$ are \textit{physically equivalent}.
Such modification of the potentials at the lattice sites implies that the ground potential is fixed as the average of the potentials is zero, that is, $ \mathbf{V}^\prime \mathbf{f}=0$.
Then, the eigenvalue of  $\mathbf{L}^{\prime} = \mathbf{L} + \mathbf{f}\circ \mathbf{f}$ 
corresponding to the eigenvector $\mathbf{f}$ is no longer zero, but rather $\mathbf{f}^2=N$.   
Thus, the modified Laplacian $\mathbf{L}^{\prime}$ is invertible, and the Green operator  
\begin{align}
	\mathbf{G} &= -{\left(\mathbf{L}^{\prime}\right)}^{-1}
	=- {\left(\mathbf{L} + \mathbf{f}\circ \mathbf{f} \right)}^{-1}
\end{align}
can now be used in equation~(\ref{Rij:eq}) 
to calculate the two-point resistance or impedance in finite networks.
The advantage of our method is that it allows one to determine the two-point resistance without calculating the eigenvalues of the Laplacian.
Note that the above procedure is related to the calculation of the Moore–Penrose generalized inverse~\cite{Xio_Gutman_2003}.

\section{Derivation of the resistance for the lake shown in
	figure~\ref{bonds_2x3hole:fig}(a)}
\label{Hole_2_missing_nodes:app}

From equations~(\ref{G_ri_rj:eq}) and (\ref{G_pq:eq})  we find that
\begin{subequations}
	\begin{align}
		\langle \alpha | \, \hat{G}_0\,  |\alpha \rangle
		&=\frac{1}{2}\, \left(
		\begin{array}{ccccc}
			1 & a & a & 0 & b \\
			a & 1 & 0 & a & c \\
			a & 0 & 1 & a & b \\
			0 & a & a & 1 & c \\
			b & c & b & c & 1 \\
		\end{array}
		\right), 
	\end{align}
	where 
	$a=2 R_0(1,1)-2R_0(1,0)= \frac{4}{\pi}-1$,
	$b=2R_0(1,0)- R_0(1,1))= 1-\frac{2}{\pi}$ and 
	$c=R_0(1,1)+R_0(2,0)-R_0(1,0)-R_0(2,1)= 2-\frac{6}{\pi}$.
	
	Using equations~(\ref{ri_G0_a:eq}), (\ref{a_G0_rj:eq}) and (\ref{gu_vg:eq}) 
	for sites $\mathbf{r}_i =\left(0,2\right)$ and 
	$\mathbf{r}_j =\left(3,0\right)$ we obtain
	\begin{align}
		\mathbf{U}^{\left(i\right)} &=	\langle i |\, \hat{G}_0\, |\alpha \rangle 
		= \frac{1}{2} \, \left(w_1,w_2,w_3,w_4,w_1 \right) , \\[2ex]
		\mathbf{V}^{\left(j\right)} &=	\langle \alpha |\, \hat{G}_0\, | j\rangle 
		=\frac{1}{2} \, {\left(w_4,w_3,w_2,w_1,w_5 \right)}^T ,
	\end{align}	
	where $w_1= R_0(1,1)-R_0(1,0),
	w_2= R_0(2,1)-R_0(2,0),
	w_3= R_0(1,1)-R_0(2,1),
	w_4= R_0(2,1)-R_0(2,2)$ and $w_5 = R_0(2,1) - R_0(3,1)$. 
	Using table~\ref{R0:table} we obtain
	\begin{align}
		\mathbf{U}^{\left(i\right)} &=	
		\frac{1}{4\pi}\,\left(4-\pi ,16-5 \pi ,\pi -4,
		\frac{8}{3}-\pi ,4-\pi \right) , \\[2ex]
		\mathbf{V}^{\left(j\right)} &=
		\frac{1}{4\pi}\, {\left(\frac{8}{3}-\pi ,
			\pi -4,16-5 \pi ,4-\pi ,7 \pi-\frac{68}{3}\right)}^T .
	\end{align}
\end{subequations}
Then the resistance can be obtained from equation~(\ref{R_ij_perturbed:eq}), and the result 
is given by equation~(\ref{R_res_fig3:eq}) in the main text. 

\section{Perturbation of triangular lattice shown in figure~\ref{6szog_missing_A:fig}(a)}
\label{3szog_hexagon:app}

\begin{figure}[hbt]
	\centering 
	\includegraphics[scale=0.5]{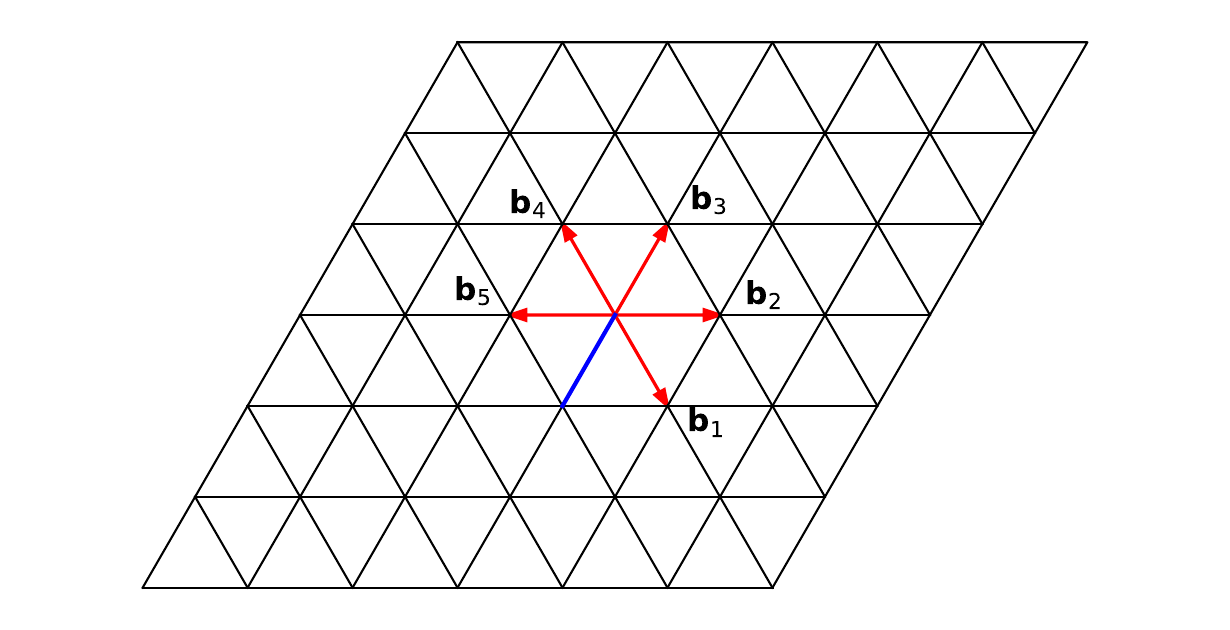}
	\caption{The bond vectors are $\mathbf{b}_1, \mathbf{b}_2, \dots, \mathbf{b}_5 $ (red arrows) corresponding to the five removed resistors and  there is 
		a single dangling bond (blue line). The origin of the coordinate system is at the center of the hexagon.
		\label{6szog_missing_B:fig}}
\end{figure}

As elucidated in section~\ref{gen_theory:sec}, the introduction of a dangling bond results in the removal of five bonds from the perfect lattice. Each removed bond is denoted by an arrow depicted 
in figure~\ref{6szog_missing_B:fig}. 
In the perfect triangular network the lattice points can be defined as $\mathbf{r} = m \mathbf{a}_1 + n \mathbf{a}_2$, where $m$ and $n$ are integers, and $\mathbf{a}_1$ and $\mathbf{a}_2$ are  vectors forming a $60^\circ$ angle between them (in the present work 
we take $\mathbf{a}_1 = \mathbf{b}_2$ and $\mathbf{a}_2 = \mathbf{b}_3$ in figure~\ref{6szog_missing_B:fig}). 
Hereafter, the lattice point $\mathbf{r}$ will be represented as $(m, n)$.

Now, the Green's operator $\hat{G}_0$ of the unperturbed resistor network corresponds to the perfect triangular lattice. 
To obtain the matrix $\mathbf{B}$ and the two vectors $\mathbf{U}^{\left(i\right)}$ and 
$\mathbf{V}^{\left(j\right)}$ in equation~(\ref{R_ij_perturbed:eq}), we need to know the resistances in the perfect triangular lattice. 
In table~\ref{R3szog:table} we give a few values of the resistance for a perfect infinite triangular resistor network~\cite{Atkinson_Steenwijk_10.1119/1.19311,sajat_perfect_AJP_10.1119/1.1285881}. 
\begin{table}[hbt]
	\caption{\label{R3szog:table}
		Resistance $R_{\triangle}(m,n)$ in an infinite triangular lattice in units of $R$.}
	\begin{indented}
		\item[]\begin{tabular}{@{}lllll}
			\br
			$m \backslash n$ &	0 & 1 & 2 & 3 \\
			\mr
			0 & 0 & $\frac{1}{3}$ & $\frac{8}{3}-\frac{4
				\sqrt{3}}{\pi }$ & $27-\frac{48 \sqrt{3}}{\pi
			}$ \\[2ex]
			1 & $\frac{1}{3}$ & $\frac{2 \sqrt{3}}{\pi
			}-\frac{2}{3}$ & $\frac{10 \sqrt{3}}{\pi }-5$
			& $\frac{128 \sqrt{3}}{\pi }-70$ \\[2ex]
			2 & $\frac{8}{3}-\frac{4 \sqrt{3}}{\pi }$ &
			$\frac{10 \sqrt{3}}{\pi }-5$ & $16-\frac{28
				\sqrt{3}}{\pi }$ & $123-\frac{222
				\sqrt{3}}{\pi }$ \\[2ex]
			3 & $27-\frac{48 \sqrt{3}}{\pi }$ & $\frac{128
				\sqrt{3}}{\pi }-70$ & $123-\frac{222
				\sqrt{3}}{\pi }$ & $\frac{3978 \sqrt{3}}{5
				\pi }-438$  \\[2ex]
			\br
		\end{tabular}
	\end{indented}
\end{table}
For negative $m,n$ the resistance can be obtained from the symmetry of the triangular lattice. 

From equations~(\ref{G_ri_rj:eq}) and (\ref{G_pq:eq})  we find that 
\begin{subequations}
	\begin{align}
		\label{3szog_Bm:eq}
		\langle \alpha | \, \hat{G}_0\,  |\alpha \rangle	&= 
		\left(
		\begin{array}{ccccc}
			\frac{1}{3} & \frac{1}{6} &
			\frac{2}{3}-\frac{\sqrt{3}}{\pi } & \frac{2
				\sqrt{3}}{\pi }-1 & \frac{2}{3}-\frac{\sqrt{3}}{\pi
			} \\
			\frac{1}{6} & \frac{1}{3} & \frac{1}{6} &
			\frac{2}{3}-\frac{\sqrt{3}}{\pi } & \frac{2
				\sqrt{3}}{\pi }-1 \\
			\frac{2}{3}-\frac{\sqrt{3}}{\pi } & \frac{1}{6} &
			\frac{1}{3} & \frac{1}{6} &
			\frac{2}{3}-\frac{\sqrt{3}}{\pi } \\
			\frac{2 \sqrt{3}}{\pi }-1 &
			\frac{2}{3}-\frac{\sqrt{3}}{\pi } & \frac{1}{6} &
			\frac{1}{3} & \frac{1}{6} \\
			\frac{2}{3}-\frac{\sqrt{3}}{\pi } & \frac{2
				\sqrt{3}}{\pi }-1 & \frac{2}{3}-\frac{\sqrt{3}}{\pi
			} & \frac{1}{6} & \frac{1}{3} \\
		\end{array}
		\right).
	\end{align}	
	Using equations~(\ref{ri_G0_a:eq}), (\ref{a_G0_rj:eq}) and (\ref{gu_vg:eq}) 
	for sites $\mathbf{r}_i =\left(1,0\right)$ and 
	$\mathbf{r}_j =\left(-1,0\right)$  we have 
	\begin{align}
		\mathbf{U}^{\left(i\right)}  &= 
		\langle i |\, \hat{G}_0\, |\alpha \rangle = 
		\left(0,\frac{1}{6},0,\frac{1}{2}-\frac{\sqrt{3}}{\pi
		},\frac{2 \sqrt{3}}{\pi }-\frac{7}{6}\right), \\[2ex]
		\mathbf{V}^{\left(j\right)}  &=
		\langle \alpha |\, \hat{G}_0\, | j\rangle =
		\left(\frac{1}{2}-\frac{\sqrt{3}}{\pi },\frac{2
			\sqrt{3}}{\pi
		}-\frac{7}{6},\frac{1}{2}-\frac{\sqrt{3}}{\pi
		},0,\frac{1}{6}	\right).
	\end{align}
\end{subequations}
Note that in these calculations the resistances $R_0(m,n)$ correspond to that of the triangular lattice, $R_\triangle(m,n)$ tabulated in table~\ref{R3szog:table}.  

Now using equation~(\ref{R_ij_perturbed:eq}) the resistance between 
sites $\mathbf{r}_i =\left(1,0\right)$ and 
$\mathbf{r}_j =\left(-1,0\right)$ can be obtained, and it is given by equation~(\ref{3szog_hexagon_missing_R:eq}) in the main text.

\section*{References}

\bibliographystyle{iopart-num}



\providecommand{\newblock}{}

\end{document}